\documentclass[reprint, twocolumn, nofootinbib, amsmath, amssymb, aps, pra, floatfix ]{revtex4-1}

\usepackage[english]{babel}
\usepackage[utf8x]{inputenc}
\usepackage[T1]{fontenc}

\usepackage{amsmath}
\usepackage{graphicx}
\usepackage[colorlinks=true, allcolors=blue]{hyperref}
\usepackage{dsfont}

\newcommand{\ctuF}{\cos(\theta_{2})}
\newcommand{\stuF}{\sin(\theta_{2})}
\newcommand{\ctdF}{\cos(\theta_{1})}
\newcommand{\stdF}{\sin(\theta_{1})}
\newcommand{\td}{\theta_{F}}
\newcommand{\tu}{\theta_{F+1}}
\newcommand{\tdF}{\theta_{1}}
\newcommand{\tuF}{\theta_{2}}
\newcommand{\Fu}{F+1}
\newcommand{\Fd}{F}

\newcommand{\omegaRF}{\omega_{\textrm{RF}}}

\newcommand{\ez}{\mathbf{e}_z}
\newcommand{\ex}{\mathbf{e}_x}
\newcommand{\ey}{\mathbf{e}_y}
\newcommand{\xboldsymbol}{\mathbf}
\newcommand{\BB}{\ensuremath{\xboldsymbol{\mathcal{B}}}}

\begin{document}

\title{Microwave spectroscopy of radio-frequency dressed $^{87}$Rb}
\author{G.A. Sinuco-Leon}
\thanks{These two authors contributed equally}
\affiliation{Department of Physics \& Astronomy, University of Sussex, Falmer, Brighton, BN1 9QH, UK}

\author{B.M. Garraway}
\affiliation{Department of Physics \& Astronomy, University of Sussex, Falmer, Brighton, BN1 9QH, UK}

\author{H. Mas}
\thanks{These two authors contributed equally}
\author{S. Pandey}
\author{G. Vasilakis}
\author{V. Bolpasi}
\author{W. von Klitzing}
\affiliation{
Institute of Electronic Structure and Laser, Foundation for Research and Technology—Hellas, Heraklion 70013, Greece
}

\author{B. Foxon}
\thanks{These two authors contributed equally }
\author{S. Jammi}
\thanks{These two authors contributed equally }
\author{K. Poulios}
\author{T. Fernholz}
\affiliation{
School of Physics \& Astronomy, University of Nottingham, University Park, Nottingham NG7 2RD, UK
}

\date{\today}
         
\begin{abstract}
We study the hyperfine spectrum of atoms of $^{87}$Rb dressed by a radio-frequency field, and present experimental results in three different situations: freely falling atoms, atoms trapped in an optical dipole trap and atoms in an adiabatic radio-frequency dressed shell trap. In all cases, we observe several resonant side bands spaced (in frequency) at intervals equal to the dressing frequency, corresponding to transitions enabled by the dressing field. We theoretically explain the main features of the microwave spectrum, using a semi-classical model in the low field limit and the Rotating Wave Approximation for alkali-like species in general and $^{87}$Rb atoms in particular. As a proof of concept, we demonstrate how the spectral signal of a dressed atomic ensemble enables an accurate determination of the dressing configuration and the probing microwave field. 
\end{abstract}

\maketitle

\section{Introduction}

The recent developments from the precise control of cold atoms \cite{ChuNobel,PhillipsNobel,CornellNobel,KetterleNobel} have paved the way to many breakthrough experimental and theoretical results \cite{arndt_njp_2012,amico_njp_2017}.  These span a range which runs from fundamental to applied physics, including quantum simulation \cite{bloch_revmodphys_2008}, atom interferometry \cite{Cronin2009,navez_njp_2016}, high precision atomic clocks \cite{treutlein_pra_2006, sarkany_pra_2014} and sensitive compact quantum sensors \cite{bohi2012simple,becker_nat_2018}.  Amongst these developments, radio-frequency (RF) and microwave (MW) dressing \cite{garraway_jphysb_2016,perrin_advamop_2017} have provided the means to generate new types of control and trapping potentials for cold atoms.  By combining magnetic fields at different frequencies, from DC to RF and MW, one can create highly non-trivial potential landscapes.  These can have complex geometries that are robust against low-frequency environmental noise \cite{treutlein_pra_2006, morizot_euphysjd_2008} and can also be transformed and manipulated adiabatically \cite{garraway_jphysb_2016, bohi_nphys_2009}.  This provides a versatile platform to investigate the physics of non-trivial topologies, e.g.\ shell potentials \cite{colombe_epl_2004}, multiple nested shell potentials \cite{PhysRevA.97.013616}, toroidal surfaces \cite{fernholz_pra_2007} and ring-shaped structures \cite{fernholz_pra_2007, sherlock_pra_2011,navez_njp_2016,Pandey2019,Lesanovsky2007PRL}. The dressed manifolds of different hyperfine states can often be coupled and manipulated independently \cite{Sinuco2016}.  This together with the robustness to temporal and spatial noise \cite{PhysRevLett.98.263201}, makes dressed potentials an ideal candidate for an interferometric, or general atomtronic, platform \cite{bohi_nphys_2009, hofferberth_nat_2007, berrada_ncomms_2013, stevenson_prl_2015, amico_njp_2017}.  However, the complexity of these potentials means that when additional fields are used to probe an atom, many new transition lines are found. This rich spectral panorama forms the subject of this paper.

We present an experimental and theoretical study of the response of RF-dressed atoms of $^{87}$Rb to MW radiation for the full range of relevant microwave frequencies.  We identify qualitatively and quantitatively how the microwave spectrum emerges from probing the RF-dressing, and observe the signatures of the spectrum in three common experimental situations.
In the following Sec.~\ref{sec:inter-an-alkali} there is a theoretical description of the internal dynamics of alkali-like atomic systems driven by one radio-frequency and one microwave field in the limit of a linear Zeeman shift and a weak RF field.
We then
present experimental results corresponding to three different scenarios: freely falling atoms (Sec.\,\ref{sec:exp-data-Nottingham}), atoms in an optical dipole trap (Sec.\,\ref{sec:exp-data-Crete}), and atoms in an RF-dressed shell trap (Sec.\,\ref{sec:atomsinashell}).  In each case, we describe the main features of the microwave spectrum and compare them with our theoretical model.  Finally, in our closing section (Sec.\,\ref{sec:conclusions}), we provide a general outlook of our findings and comment on future applications.

\section{Interaction of an alkali atom with radio-frequency and microwave magnetic fields}
\label{sec:inter-an-alkali}

The internal dynamics of an alkali atom in its electronic ground state interacting with a weak, time-dependent magnetic field $\xboldsymbol{B}(t)$ are governed by the 
Hamiltonian:
\begin{equation}
\hat{H} = \frac{A}{\hbar^2} \hat{\mathbf{I}}\cdot\hat{\mathbf{J}} + \frac{\mu_\text{B}}{\hbar} (g_I \hat{\mathbf{I}} + g_J \hat{\mathbf{J}}) \cdot \xboldsymbol{B}(t) ,
\label{eq:H_generic}
\end{equation}
where $A$ is a hyperfine structure constant, and $\mu_\text{B}$ is the Bohr magnetron. The factors
$g_I$ and $g_J$ are the nuclear and electronic $g$-factors, respectively They have the corresponding angular momentum operators $\hat{\mathbf{I}}$ and $\hat{\mathbf{J}}$.

Here we consider a magnetic field with three contributions: a time-independent (DC) part and two harmonically oscillating components at radio-frequency (RF) and microwave (MW) frequencies:
\begin{equation}
  \xboldsymbol{B}(t) = B_{\textrm{DC}} \, \ez   
  +  \xboldsymbol{B}_{\textrm{RF}}(t) + \xboldsymbol{B}_{\textrm{MW}}(t) .
\label{eq:TotalField}
\end{equation}
Without loss of generality, we choose a quantization axis (unit vector $\ez$) along the direction of the static field of strength $B_{\textrm{DC}}$.

For zero external magnetic field, the coupling between the nuclear and electronic magnetic moments (with quantum numbers $I$ and $J=L\pm S$) defines two hyperfine manifolds with different total angular momentum and corresponding quantum number $F=I \pm J$, which are split by an energy gap of $\Delta E_{\text{hfs}} = AJ(2I+1)$.  The static component of the field, $B_{\textrm{DC}}$, lifts the degeneracy within each hyperfine manifold (Zeeman splitting).  When the hyperfine splitting is much larger than the energy associated with the applied magnetic fields, that is, $\Delta E_{\text{hfs}} \gg \mu_\text{B}(B^2_{\text{DC}}+|\xboldsymbol{B}_{\textrm{RF}}|^2+ |\xboldsymbol{B}_{\text{MW}}|^2)^{1/2}$, the total angular momentum $F$ remains a good quantum number, and the atomic spectrum can be conveniently described in the basis $\{\left|F=I+J,m_\text{F}\right\rangle \} \oplus \left|F=I-J,m_\text{F} \right\rangle\}$, with $m_\text{F}={-F,...,F}$ \cite{mischuck_pra_2012}.

In this basis, the static part of the Hamiltonian Eq.~(\ref{eq:H_generic}) can be linearly approximated as
\begin{equation}
	\hat{H}_0 = \sum_\text{F} \left(E_F + \mu_\text{B} g_F \hat{F}_zB_\text{DC}\right)\hat{\mathds{1}}_F , 
    \label{eq:H_0}
\end{equation}
where we have defined partial identity operators to project onto the hyperfine manifolds,
\[\hat{\mathds{1}}_F=\sum_{m_\text{F}=-F}^{F}|F,m_\text{F}\rangle\langle F,m_\text{F}|, \]
and we have used the property $[\hat{\mathds{1}}_F,\hat{F}_z]=0$.
Energies and $g_F$-factors for the two manifolds are given by 
\begin{equation}
E_F = \frac{1}{2}A \left(F(F+1)-I(I+1)-J(J+1)\right),
\end{equation}
and
\begin{eqnarray}
g_F &=&  g_J \frac{F(F+1) + J(J+1) - I(I+1)}{2F(F+1)}  \nonumber \\
& &+ g_I \frac{F(F+1) - J(J+1) + I(I+1)}{2F(F+1)} ,
\label{eq:gFs}
\end{eqnarray}
(e.g.\ see \cite{steck_data_2015} and Refs.\ [17,20,25] therein).

\begin{figure}[!htb]
\centering
\includegraphics[width=\columnwidth]{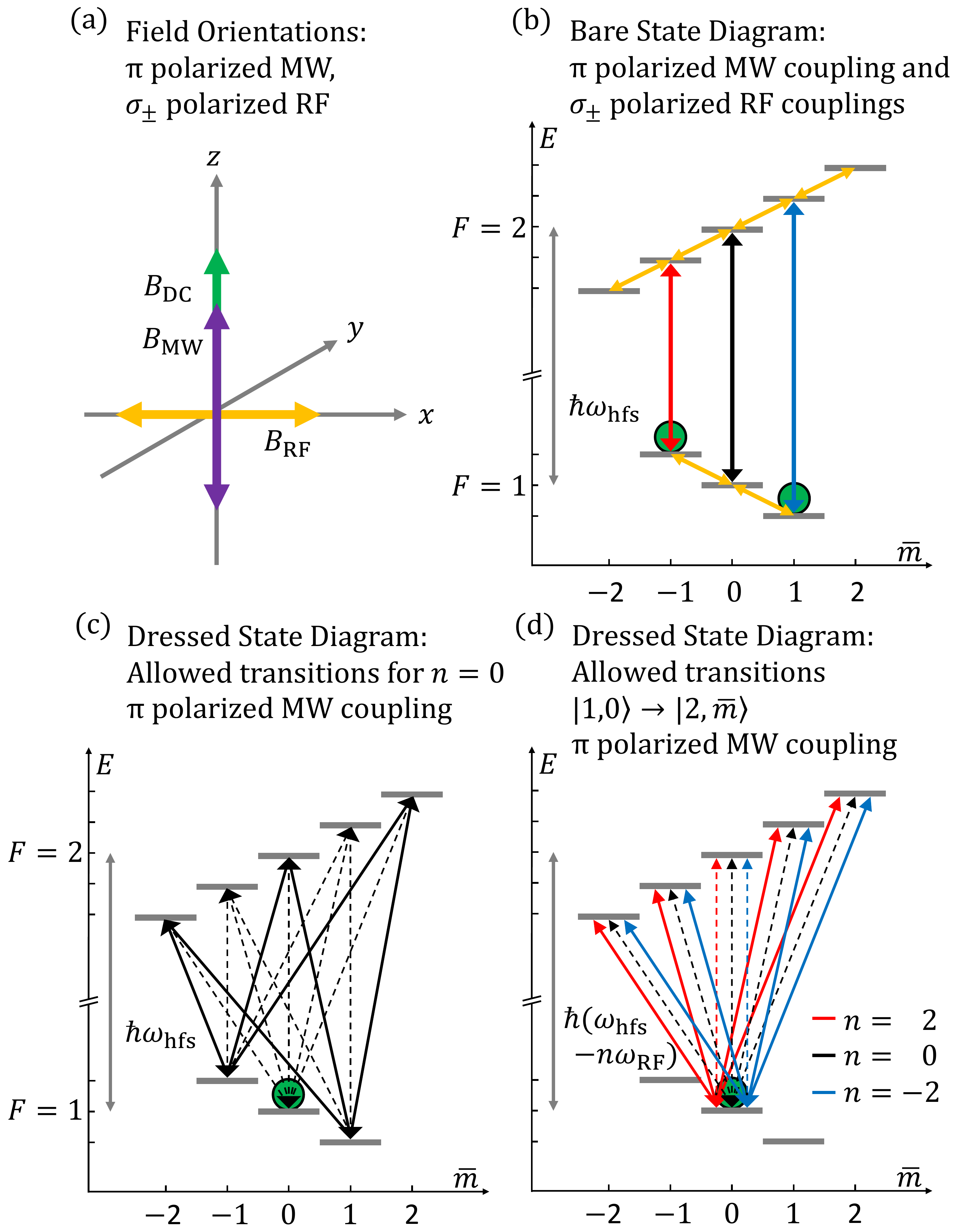}
\caption{\label{fig:Levels87Rb}%
Energy level scheme and couplings for the RF dressed, electronic ground state of $^{87}$Rb with two hyperfine manifolds of total angular momentum $F=1,2$. (a) The figure shows a linearly polarised RF field, orthogonal to the static field, and the example of a $\pi$-polarized MW field, i.e.\ the magnetic field oscillating parallel to the static field. (b) In the laboratory frame, the RF (orange) and MW (red, black, blue) fields with frequencies $\omega_{\textrm{RF}}$ and $\omega_{\textrm{MW}}$ induce 
intra- and inter-manifold couplings, respectively.  The example shows the resonantly dressed $|1,\bar{m}=0\rangle$ state, which is a superposition of the two bare states $|1,-1\rangle$ and $|1,+1\rangle$, as marked by the green circles. In the dressed picture, the RF field becomes a component of an effective static field, and the MW field can in principle couple any pair of states from the two manifolds. In the ideal case of $g_1=-g_2$ and on RF resonance some transitions are forbidden, indicated by dashed lines, which is shown in (c) for driving near the zero field hyperfine splitting frequency (n=0). Here, the dressed state $|1,\bar{m}=0\rangle$ is not coupled. In (d) the situation is shown for only the state $|1,\bar{m}=0\rangle$ but for all apparent sidebands $\omega_{\textrm{MW}} + n \omega_{\textrm{RF}}$, with $n=+2,0,-2$ (red, black, blue), which are resonant for frequencies $\omega_{\textrm{MW}}$ near the three corresponding $\pi$-transitions between bare states.   
See Appendix \ref{sec:Appendix3}.}
\end{figure}

The arrangement of energy levels and coupling fields is illustrated for the $^{87}$Rb ground state in Figure~\ref{fig:Levels87Rb}
for the example of a $\pi$-polarized MW field. In the case of  $^{87}$Rb ($I=3/2$), the two $g_F$-factors given by Eq.~(\ref{eq:gFs}) are $g_1 = -0.50182671$ for the lower manifold
and $g_2 = 0.49983642$ for the upper manifold.

The two time-dependent terms in Eq.\,(\ref{eq:TotalField}) oscillate at frequencies close to the resonance condition and cause two different types of transitions
written as
$\left|F',m_\text{F}' \right\rangle \leftrightarrow \left| F,m_\text{F}\right\rangle$. The radio-frequency field, $\xboldsymbol{B}_{\textrm{RF}}(t)$ oscillates at a frequency of the order of the Zeeman splitting,  $\omega_{\textrm{RF}} \sim |g_F|\mu_\text{B} B_{\textrm{DC}}/\hbar$, which is typically in the range of $10$s kHz -- MHz. It is convenient to represent the corresponding atom-field interaction term in the basis of total angular momentum ${\left|  F, m_\text{F}\right\rangle}$. In this basis, the hyperfine interaction $A \hat{\mathbf{I}}\cdot \hat{\mathbf{J}}$ splits the energy spectrum in two blocks of Zeeman sub-states of total angular momentum $F=|I\pm J|$,
and transitions within each block, corresponding to $|F-F'| = 0$, are near-resonantly coupled by the RF field. This part of the Hamiltonian can be approximated by a term $\hat{H}_\textrm{RF}\propto g_F\xboldsymbol{B}_\textrm{RF}\cdot\hat{\mathbf{F}}$. 

The microwave field $\xboldsymbol{B}_{\text{MW}}(t)$ oscillates at a frequency of the order of the hyperfine splitting: $\omega_{\text{MW}} \sim |E_{I+J}-E_{|I-J|}|/\hbar$, which for alkali atoms ranges between $0.2$ - $10$~GHz \cite{PhysRevA.96.062502, steck_data_2015}. In this case, the couplings between blocks of states defined by the hyperfine coupling are resonant and the MW field leads to transitions between states belonging to different hyperfine manifolds such that $|F-F'| = 1$. For this part of the Hamiltonian, we neglect the small nuclear magnetic moment due to $g_I\ll g_J$, and approximate it by a term of the form $\hat{H}_\text{MW}\propto g_J\xboldsymbol{B}_\text{MW}\cdot\hat{\mathbf{J}}$, see below. 

The two oscillating fields can be expressed in spherical polarization components defined with respect to the direction of the static field \cite{perrin_advamop_2017} as
\begin{align}
  \xboldsymbol{B}_{\textrm{AC}}(t)  &=
  \BB_{\textrm{AC}}
  e^{-i\omega_{\textrm{AC}} t}
+  \BB^{\ast}_{\textrm{AC}}
e^{i\omega_{\textrm{AC}} t} ,
\intertext{with complex amplitudes}
  \BB_{\textrm{AC}}
 &= B_{\textrm{AC},+} \xboldsymbol{e}_{+} + B_{\textrm{AC},-}
   \xboldsymbol{e}_{-} + B_{\textrm{AC},0}\xboldsymbol{e}_0
   \,.
\end{align}
Here we let $\textrm{AC}\rightarrow \textrm{RF,MW}$  and we have used the definitions \cite{perrin_advamop_2017}
\begin{align}
\xboldsymbol{e}_0 &= \ez,  &B_{\textrm{AC},0} &= \frac{B_{\textrm{AC,z}}e^{-i \phi_z}}{2}\nonumber \\
\xboldsymbol{e}_\pm &= \mp \frac{ \ex \pm i \ey }{\sqrt{2}}, & B_{\textrm{AC},\pm} &= \frac{
     \mp B_{\textrm{AC},x}  e^{-i\phi_x}
     + i B_{\textrm{AC},y}  e^{-i\phi_y}   }{2\sqrt{2}} ,
     \label{eq:PolarComponentsDefinition}
\end{align}
where $\phi_i$ represent the phases of the $i$th component of the AC field.
Using this parametrisation of the fields and taking into account the range of frequencies of each component, the RF and MW interaction Hamiltonians are given by:
\begin{widetext}
\begin{eqnarray}
\hat{H}_{\textrm{RF}}(t) &=& \sum_\textrm{F} \hat{\mathds{1}}_F \sum_{\sigma\in\{+,-,0\}} \eta_{\sigma} \mu_\text{B} g_F \left( B_{\textrm{RF},\sigma} e^{-i \omega_{\textrm{RF}}t} \hat{F}_{\sigma} + B_{\textrm{RF},\sigma}^* e^{i \omega_{\textrm{RF}}t} \hat{F}_{-\sigma}\right) ,
\label{eq:H_RF} \\
\hat{H}_{\text{MW}}(t) &=& \sum_{\sigma\in\{+,-,0\}}  \eta_{\sigma} \mu_\text{B} g_J \left( B_{\textrm{MW},\sigma} e^{-i \omega_{\textrm{MW}}t} \hat{J}_{\sigma} +  B_{\textrm{MW},\sigma}^* e^{i \omega_{\textrm{MW}}t} \hat{J}_{-\sigma}\right) ,
\label{eq:H_MW}
\end{eqnarray}
\end{widetext}
where the raising and lowering angular momentum operators are defined by $\hat{F}_\pm = (\hat{F}_x \pm i \hat{F}_y)$, with similar expressions for the electronic angular momentum $\hat{J}_\pm$. The factors $\eta_{+1}=-1/\sqrt{2}$, $\eta_{-1}= 1/\sqrt{2}$, $\eta_0=1$ follow from our definitions in Eq.~(\ref{eq:PolarComponentsDefinition}).

In the next section,  we describe how the Rotating Wave Approximation (RWA) leads to an approximate description of the internal dynamics of alkali atoms subjected to this bi-chromatic field.

\subsection{RF-dressing in the Rotating Wave Approximation}
\label{sec:RWA}

Let us first consider the case where there is no microwave field, i.e.\ $\xboldsymbol{B_{\text{MW}}}(t) =0$. Then the Hamiltonian becomes,
\begin{equation}
\hat{H} = \hat{H}_0 + \hat{H}_\textrm{RF}(t) ,
\label{eq:H_DCRF}
\end{equation}
where $\hat{H}_0$ is defined in Eq.\,(\ref{eq:H_0}) and $\hat{H}_\textrm{RF}$ is given by Eq.\,(\ref{eq:H_RF}).
The resulting dynamics can be described in the dressed basis, i.e.\ by moving to a rotating frame where the most relevant component of the field becomes time-independent, 
and diagonalisation of the resulting Hamiltonian becomes analytically tractable. More specifically, we describe the driven atom in the rotating frame of reference that follows from the unitary transformation
\begin{eqnarray}
\hat{U}_z(\omega_{\textrm{RF}}t) &=&\sum_\textrm{F} \hat{\mathds{1}}_F \exp\left[-i\text{sgn}(g_F)\omega_\textrm{RF}t \hat{F}_z\right],
\label{eq:RotationZ}
\end{eqnarray}
which corresponds to geometric rotations about the $z$-axis at frequency $\omega_\textrm{RF}$, but in opposite directions due to the opposite sign of the $g_F$ factors.
In the rotating frame, the Hamiltonian (\ref{eq:H_DCRF}) becomes: 
\begin{eqnarray}
\hat{\tilde{H}}  &=& \hat{U}_z^\dagger \hat{H} \hat{U}_z - i \hbar \hat{U}_z^\dagger \partial_t \hat{U}_z \nonumber \\
&\approx& \sum_\text{F} \hat{\mathds{1}}_F \left[  E_F +  (\mu_\text{B} g_F B_{\textrm{DC},z} -    \text{sgn}(g_F) \hbar\omega_{\textrm{RF}}) \hat{F}_z  \right. \nonumber \\
& &\left. +  \frac{\mu_\text{B} g_F}{2} \sqrt{2} (B_{\textrm{RF},\text{sgn}(g_F)} \hat{F}_{+} + B_{\textrm{RF},\text{sgn}(g_F)}^*\hat{F}_{-}) \right] ,
\nonumber\\&&\label{eq:H_RF_Rot}
\end{eqnarray}
where we have neglected inter-manifold couplings and applied the Rotating Wave Approximation (RWA), which consists of neglecting time-dependent terms oscillating at angular frequency $\pm2\omega_\textrm{RF}$. This procedure is valid as long as the processes associated with these terms are far from being resonant. The RF-dressed states are defined as the eigenstates of Eq.\,(\ref{eq:H_RF_Rot}), which can be obtained by performing a second (time-independent) rotation within each hyperfine manifold,
\begin{eqnarray}
\hat{U}_y &=& \sum_\text{F} \hat{\mathds{1}}_F \exp\left( -i\theta_F \hat{F}_y\right) ,
\label{eq:RotationY}
\end{eqnarray}
where
\begin{equation}
\theta_{F} = \frac{\pi}{2} - \tan^{-1} \left( \frac{ B_{\textrm{DC}} -  \hbar \omega_{\textrm{RF}}/(\mu_B |g_F|)}{\sqrt{2} B_{\textrm{RF},\textrm{sgn}(g_F)}} \right) .
\end{equation}
The resonance condition $\mu_\text{B} |g_F| B_{DC}= \hbar \omega_\textrm{RF}$ depends on $F$ and is shifted by $(g_2+g_1)\mu_\text{B} B_{\text{DC}}
$  (or $h\cdot 2.78666$~kHz/G in the case of  $^{87}$Rb),
which causes a small difference in the shape of the dressed potentials, as we will see below.

In the basis of RF dressed states, the Hamiltonian $\hat{\bar{H}} = \hat{U}_y^\dagger \hat{\tilde{H}} \hat{U}_y$ becomes
\begin{eqnarray}
\hat{\bar{H}}= \sum_\text{F}  \hat{\mathds{1}}_F \left[ E_F + \hbar \Omega_{\textrm{RF}}^F \hat{F}_z \right] , 
\label{eq:H_RFdressed}
\end{eqnarray}
with the Rabi frequencies, $\Omega_{\textrm{RF}}^F$, defined by:
\begin{equation}
\hbar \Omega_\textrm{RF}^{F} = \mu_\text{B} g_{F} \sqrt{\left( B_{\textrm{DC}} - \frac{g_F}{|g_F|}\frac{ \hbar \omega_{\textrm{RF}}}{\mu_\text{B} g_{F}} \right)^2 + 2\left|  B_{\textrm{RF},\text{\tiny{sgn}}(g_F)} \right|^2}.
\label{eq:DressedRabiFrequency}
\end{equation}
With this construction, the dressed states are defined as a time-dependent superposition of Zeeman states, i.e.\ they can be expressed in the bare basis as:
\begin{equation}
\left| F, \bar{m}\right\rangle = \sum_{m=-F \ldots F}  e^{-i\text{sgn}(g_F) \bar{m} \omega_\textrm{RF}t } d^{F}_{m \bar{m}}(\theta_F)\left| F,m \right\rangle ,
\label{eq:dressedbare}
\end{equation}
where $d^{F}_{m\bar{m}}(\theta)$ is the Wigner $d$-matrix, 
\begin{equation}
d^{F}_{m',m}(\theta) = \left\langle F, m' \right| e^{-i\theta \hat{F}_y} \left| F,m\right\rangle ,
\label{eq:wignersmalldmatrix}
\end{equation}
which represents the rotation of the
operator $\hat{U}_y(\theta)$.
In the case of $^{87}$Rb,
the nuclear angular momentum $I=3/2$ implies that the ground state manifold splits into two hyperfine manifolds of total angular momentum $F=1,2$, with Hilbert space dimensions $3$ and $5$, respectively. Values for the $d^{F}_{m',m}(\theta)$ for rotations about the $y$-axis are presented in matrix form in Appendix \ref{app:dmatrices}. In combination with time-dependent factors in Eq.\,(\ref{eq:dressedbare}), matrices (\ref{eq:d1}) and (\ref{eq:d2}) give us the time-dependent relation between the bare and dressed representations.

When dealing with problems restricted to a single hyperfine manifold a simpler treatment is possible \cite{perrin_advamop_2017}. The unitary transformation to the basis of RF dressed states can then be expressed in terms of separate spatial rotational matrices, exploiting the equivalence between spin and spatial rotations for interactions of the form $\hat{V}= \boldsymbol{\mu}\cdot \hat{\mathbf{F}}$. More concretely, in a rotating frame reached by the unitary transformation $\hat{U}=\exp(-\theta \hat{n} \cdot \hat{\mathbf{F}})$, the interaction can be obtained using the Baker-Campbell-Hausdorff Lemma:
\begin{eqnarray}
\hat{U}^\dagger \hat{V} \hat{U} &=& \boldsymbol{\mu}\cdot \hat{U}^\dagger \hat{\mathbf{F}} \hat{U} \nonumber \\
 &=& R_{\hat{n}}(-\theta)\boldsymbol{\mu}\cdot \hat{\mathbf{F}}
\end{eqnarray}
where $R_{\hat{n}}(-\theta)$ is a $3\times 3$ matrix corresponding to the rotation by an angle $-\theta$ around the axis aligned in the direction of $\hat{n}$ \cite{fernholz_pra_2007, jammi_pra_2018}. 
Here, we are concerned with couplings between RF dressed manifolds with \textit{different} total angular momentum and therefore it is more convenient to use the transformation between the Zeeman and dressed bases as given by Eqs.~(\ref{eq:dressedbare})-(\ref{eq:wignersmalldmatrix}).

\subsection{MW coupling of RF-dressed states in the Rotating Wave Approximation}
\label{sec:mw-coupling-rf}
RF-dressed states of the electronic ground state of an alkali atom can be prepared by starting in bare states and adiabatically tuning into resonance with the dressing field. The resonance frequency is given by the Zeeman splitting, which corresponds to $\omega_{\text{RF}} \sim 2\pi\times 0.70~$kHz per Gauss for $^{87}$Rb. In this section, we study how a coherent superposition of RF-dressed states of the two hyperfine manifolds can be prepared by a applying a second field with a frequency set by the hyperfine splitting, which corresponds to $\omega_{\text{MW}} \sim 2\pi\times6.834~$GHz for $^{87}$Rb. 

This problem can be studied in the context of the response of continuously driven quantum systems, which has been the subject of theoretical and experimental study over several decades \cite{jammi_pra_2018, autler_physrev_1955, series_physrep_1978}. The experimental observations of the spectrum of off-resonant RF-dressed states made by Haroche and Cohen-Tannoudji can be understood using perturbative expansions of driven two-level systems (TLS) \cite{haroche_prl_1970, cohen-tannoudji_cras_1966, haroche_jphys_1969, zanon-willette_prl_2012, beaufils_pra_2008}. 
In addition, more recent experiments demonstrate that the modified response of resonantly RF-dressed alkali atoms to MW fields enables the encoding and manipulation of qudits exploiting the full complexity of the hyperfine manifold \cite{mischuck_pra_2012, smith_prl_2013}, and going beyond the TLS paradigm. In this section we explain how the response of RF-dressed $^{87}$Rb to a MW field can be obtained by applying a second rotating wave approximation (for the MW field), which allows us to calculate selection rules, resonant conditions and coupling strengths.

Similar to the RF case, the interaction with the MW field has contributions from both the nuclear and electronic magnetic moments. However, since the nuclear gyromagnetic factor ($g_I=-0.000995$) is three orders of magnitude smaller than the electronic one ($g_J = 2.002319$), within the RWA it is sufficient to consider only the electronic coupling in Eq.\,(\ref{eq:H_MW}).  When the atoms are continuously dressed by an RF field, the microwave field induces transitions between the dressed states defined by Eq.\,(\ref{eq:dressedbare}), which can be obtained by expressing the interaction $\hat H_{\text{MW}}$ in the dressed basis. Explicitly, this calculation corresponds to finding \cite{Sinuco2016}
\begin{equation}
\hat{\bar{H}}^\sigma_\text{MW} =  \hat{U}_y^\dagger(\tu,\td)\hat{U}_z^\dagger(t) \hat{H}^\sigma_\text{MW} \hat{U}_z(t) \hat{U}_y(\tu,\td) ,
\label{eq:transformations}
\end{equation}
where $\hat H^\sigma_{MW}$ is the contribution of the field component with polarization $\sigma$ to the MW interaction Eq.\,(\ref{eq:H_MW}), and the rotations are defined for each of the hyperfine manifolds.
After some algebraic manipulation (see Appendix~\ref{sec:matrix-elements-detail}), the matrix elements of the MW coupling are given in the dressed basis by
\begin{widetext}
\begin{eqnarray}
\left\langle F+1,\bar{m}'\right| \hat{\bar{H}}^\sigma_{\text{MW}}\left|F,\bar{m} \right\rangle &=& \eta_\sigma \mu_\text{B} g_J \sqrt{\frac{2I(I+1)}{2I+1}} \sum_{\ell=-1}^1  B^{\sigma,\ell}_{\textrm{MW}}(t) \nonumber \\
&&\times\sum_{m=-F}^{F} e^{i \omega_{\textrm{RF}}t(2m+\ell))} \times d^{F+1}_{\bar{m}',m+\ell}(-\theta^{F+1})d^{F}_{m,\bar{m}}(\theta_F)  \nonumber \\
&& \times (-1)^{(F+1-m-\ell)}
\begin{pmatrix}
F+1 & 1 & F \\
-(m + \ell)& \ell & m
\end{pmatrix}  ,
\label{eq:MWCouplingofRFdressedStates}
\end{eqnarray}
\end{widetext}
where $\eta_{+1}=-1/\sqrt{2}$, $\eta_{-1}=1/\sqrt{2}$, $\eta_0 = 1$, and
with the standard notation for the 3-j Wigner coefficients. We also use
the Wigner $d$-matrix defined in Eq.\,(\ref{eq:wignersmalldmatrix}),
and the definition
\begin{widetext}
\begin{equation}
B_{\textrm{MW}}^{\ell,\sigma}(t) = \left(B_{\textrm{MW},\sigma}\left(\frac{1+\sigma\ell}{2}\right) + B_{\textrm{MW},\sigma}^*\left(\frac{1-\sigma\ell}{2}\right) \right) e^{-i\sigma\ell\omega_{\textrm{MW}}t} + (1-|\sigma|)\delta_{\ell,0}\left(B_{\textrm{MW},\sigma} e^{-i\omega_{\textrm{MW}}t} + B_{\textrm{MW},\sigma}^*  e^{i\omega_{\textrm{MW}}t} \right) .
\label{eq:B_MW_sigmaell}
\end{equation}
\end{widetext}

Due to the transformation to the (counter) rotating frame(s), a single frequency microwave field will appear modulated, which gives rise to fictitious sidebands. According to Eq.\,(\ref{eq:MWCouplingofRFdressedStates}) the MW driving between dressed states causes coupling terms with angular frequencies equal to $\omega_{\text{MW}}$ plus multiples of the RF dressing frequency, $n\omega_{\textrm{RF}}$.
This lets us split the interaction into contributions from each MW polarization ($\sigma$) at different frequencies in the form \cite{Sinuco2016}
\begin{equation}
\hat{\bar{H}}_\textrm{MW}  = \sum_{n,\sigma} \hat{\bar{H}}^{\sigma,n}_{\textrm{MW}} e^{-i(\omega_{\textrm{MW}}+n\omega_\textrm{RF})t} + \hat{\bar{H}}^{\sigma, n \dagger}_{\textrm{MW}} e^{i(\omega_{\textrm{MW}}+n\omega_{\textrm{RF}})t}   ,
\label{eq:H_MW_dec}
\end{equation}
with $n\in[-2I,2I]$, $\sigma\in[-1,1]$ and the matrix elements defined by Eqs.~(\ref{eq:MWCouplingofRFdressedStates},\ref{eq:B_MW_sigmaell}).

The coefficients $B_{\textrm{MW}}^{\ell,\sigma}$ defined in Eq.\,(\ref{eq:B_MW_sigmaell}) lead to several relations between the matrix elements that depend on the polarization of the MW field but not on the RF dressing configuration. They give rise to a structure that reproduces the bare microwave spectrum. The $\pi$-polarised component of the MW field enables coupling at even sidebands, i.e.\ for oscillatory terms of MW frequency plus even multiples of $\omega_\textrm{RF}$. Similarly, the $\sigma_\pm$-polarised components enable coupling at the MW frequency plus odd multiples of $\omega_\textrm{RF}$, but not for the respective extremal $\omega_\textrm{MW} \pm (2\Fd +1)\omega_\textrm{RF}$. (Note that an apparent positive sideband allows for red-detuned driving in the laboratory frame.)

In general, the coupling between dressed states depends on the RF dressing configuration via the Wigner $d$-matrices. However, in agreement with symmetry considerations and conservation of the angular momentum of the atom plus radiation system, the matrix elements of each contribution to Eq.\,(\ref{eq:H_MW_dec}) satisfy the relation:
\begin{eqnarray}
\left\langle F+1,\bar{m}'\right| \hat{\bar{H}}^{\sigma,n}_\text{MW} \left|F,\bar{m} \right\rangle &=& \frac{ B_{\text{MW},\sigma}}{B_{\text{MW},-\sigma}} (-1)^{\sigma+\bar{m}'-\bar{m}-1} \nonumber \\
&& \times \left\langle F+1,-\bar{m}'\right| \hat{\bar{H}}^{-\sigma,-n}_\text{MW} \left|F,-\bar{m} \right\rangle. \nonumber \\
\label{eq:relations}
\end{eqnarray}

The MW couplings in the RF dressing configuration must meet the resonance conditions
\begin{equation}
\omega_{\textrm{MW}} + n \omega_\textrm{RF} = \omega_{\textrm{hfs}} +  \bar{m}\Omega_{\textrm{RF}}^F + \bar{m}' \Omega_{\textrm{RF}}^{F+1} , 
\label{eq:ResonantCondition}
\end{equation}
with $n\in[-2(F+1)-1,2F+1]$ and $\bar{m},\bar{m}'\in Z$ and $\Omega_{\textrm{RF}}^F$ defined in Eq.\,(\ref{eq:DressedRabiFrequency}). On the left hand side of Eq.\,(\ref{eq:ResonantCondition}) we have the oscillating frequency of the MW field observed in the dressed frame of reference, while on the right hand side we have written the quasi-energy difference between pairs of dressed states \{$\left| F \bar{m}\right\rangle$, $\left|F+1,\bar{m}' \right\rangle$\}.

\begin{figure}[!htb]
\centering
\includegraphics[width=\columnwidth]{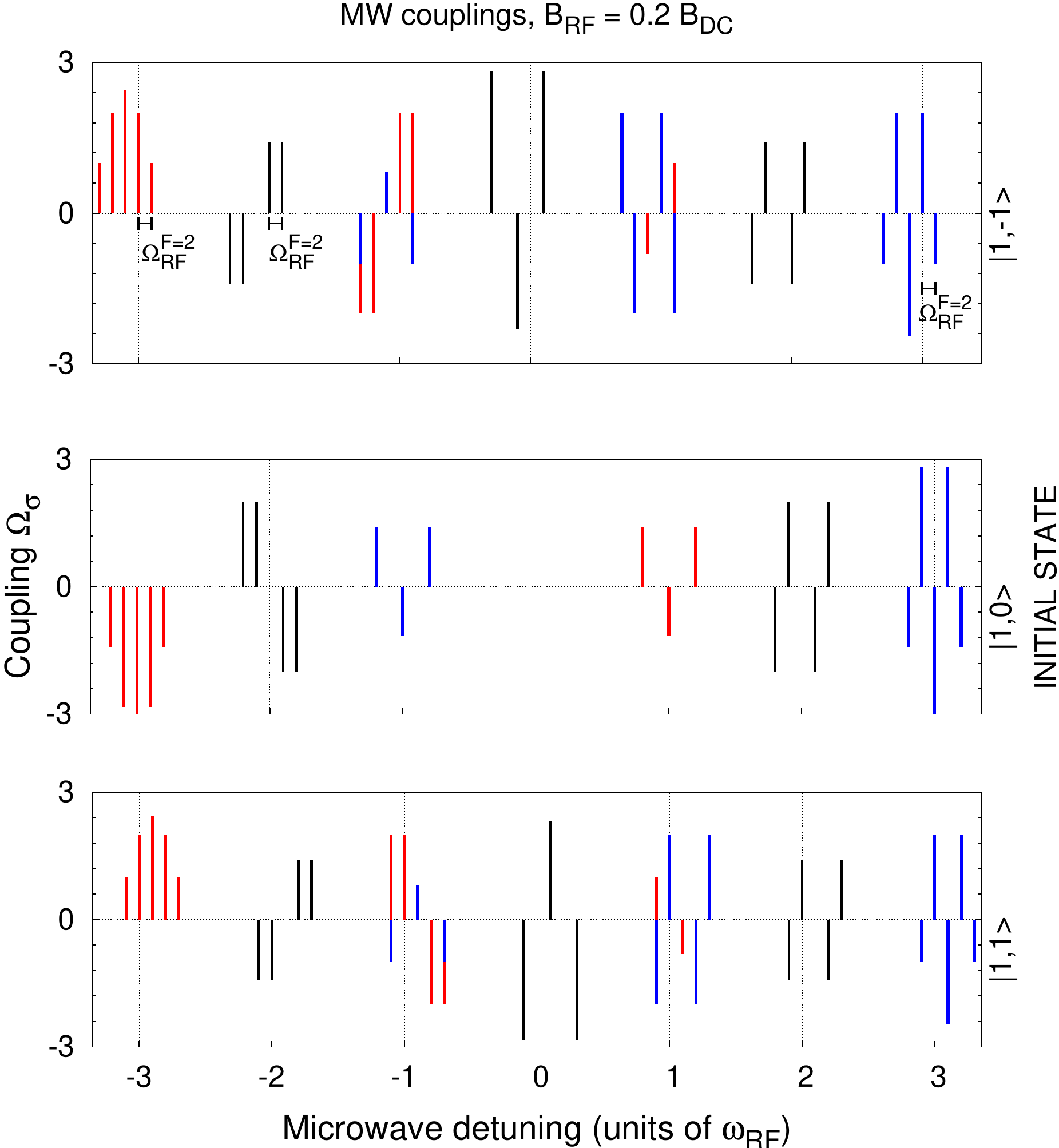}
\caption{%
\label{fig:TypicalSpectra87Rb}{Resonant frequencies and MW couplings between resonantly RF-dressed states of $^{87}$Rb calculated using Eq.~(\ref{eq:MWCouplingofRFdressedStates}) (see Appendix \ref{sec:Appendix3}). Resonances cluster around integer multiples of the RF frequency. The clusters can be associated with spherical polarization components of the microwave field (red $\sigma_-$, blue $\sigma_+$ and black $\pi$). The microwave couplings $\Omega_\sigma$ are scaled to units of $\frac{1}{16}\sqrt{\frac{3}{2}} |\eta_\sigma|\mu_\text{B} g_J B_{\text{MW},\sigma} / \hbar$. 
The horizontal axis indicates the microwave detuning from the zero field hyperfine splitting in units of the RF frequency. In this case, the dressing field is linearly polarized and orthogonal to the static field. Its amplitude is $B_\textrm{RF}=0.2 B_\text{DC}$ and the angular frequency is resonant with the Zeeman splitting $\omega_{\textrm{RF}} = \mu_\text{B} |g_F| B_\text{DC} / \hbar$, neglecting the difference between gyromagnetic factors. The $\sigma_\pm$ polarizations of the MW field are defined with phases $\phi_x = 0$ and $\phi_y = \mp\pi/2$ and $B_x = B_y > 0$.}}
\end{figure}

In Figure \ref{fig:TypicalSpectra87Rb} we depict schematically the MW spectrum of resonantly RF-dressed  $^{87}$Rb, considering as the initial state each one of the dressed sub-levels of the lower hyperfine manifold $\Fd =1$, and the three possible MW polarizations. In this case, there are $105$ potential transition frequencies corresponding to $3\times 5 = 15$ different pairs of states in the lower and upper hyperfine manifolds, coupled by terms oscillating at the $7$ different frequencies $\omega_\text{MW} + n\omega_\textrm{RF}$ with $n\in-3,3$. Resonant frequencies are given by Eq.\,(\ref{eq:ResonantCondition}) and the MW couplings are calculated with Eq.\,(\ref{eq:MWCouplingofRFdressedStates}), considering resonant RF-dressing and neglecting the difference between gyromagnetic factors. An explicit form of the couplings for $^{87}$Rb is presented in extended form in Appendix \ref{sec:Appendix3}.

Groups of resonant transitions between RF-dressed states can be labelled by the integer multiplier $n$ of the RF angular frequency in the resonant condition Eq.\,(\ref{eq:ResonantCondition}). As a consequence of the conservation of angular momentum (see Eq.\ (\ref{eq:AngularConservation}) in Appendix \ref{sec:Appendix3}), transitions in the even and odd groups are induced by $\pi$- and $\sigma_{\pm}$-polarised MW radiation, which is reminiscent of the MW transitions of bare atoms.

The analysis presented above applies to the electronic ground state of alkali-metal atoms and alkali-metal-like ions \cite{PhysRevA.96.062502}, with the total number of possible transitions and groups defined by the nuclear total spin (and then the ranges of $\bar{m},\bar{m}'$ and $n$ in Eq.\ (\ref{eq:ResonantCondition})). For instance, the MW spectrum of the RF-dressed bosonic species $^{87}$Rb, $^{39}$K, $^{23}$Na and $^{7}$Li, present the same number of resonances since the ground state of all of them is split in the manifold $F=2$ and $F=1$, though the resonant frequencies are determined by their fine and hyperfine constants.

\section{Microwave spectroscopy of RF-dressed rubidium-87}
\label{sec:exp-data}

\begin{figure}[!htb]
\centering
\includegraphics[width=0.95\columnwidth]{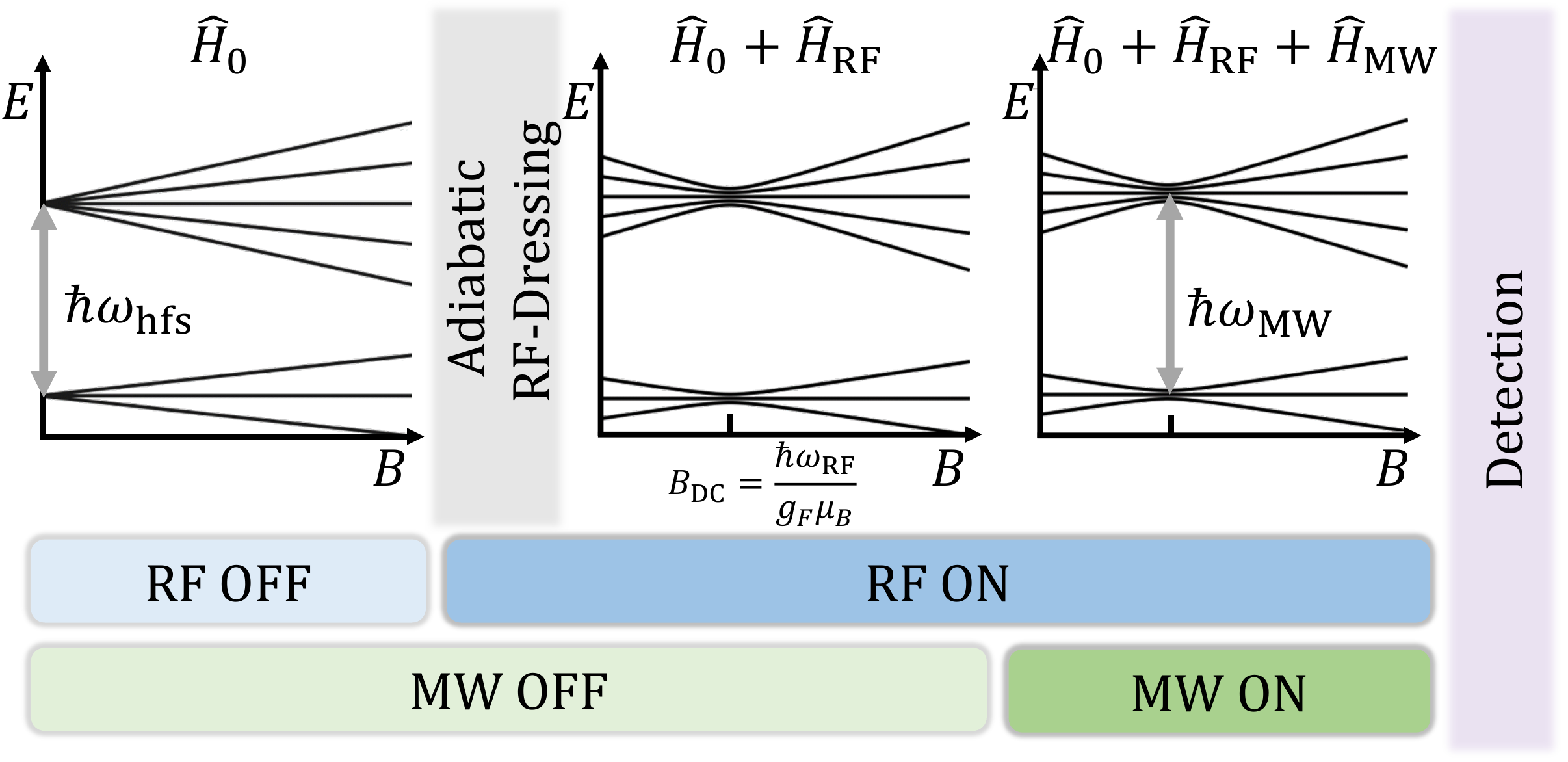} 
\caption{\label{fig:SchematicTimeSequence}%
Scheme of a generic experimental sequence. First the sample is prepared in the bare state basis, defined by $\hat{H}_{0}$. Then an RF-field is switched on and the atoms are adiabatically dressed. At this stage the total Hamiltonian is $\hat{H}=\hat{H}_{0}+\hat{H}_{\textrm{RF}}$. Afterwards, a MW-field is switched on for a short time which couples the two hyperfine manifolds via $\hat{H}=\hat{H}_{0}+\hat{H}_{\textrm{RF}}+\hat{H}_{\textrm{MW}}$. Finally, we measure the different sub-level populations.}
\end{figure}

A typical experimental sequence describing the general outline for all three experimental scenarios presented in this section is shown in Fig.\,\ref{fig:SchematicTimeSequence}, with the eigenenergies of the $^{87}$Rb hyperfine sub-levels at different stages of the sequence. We  first examine the MW spectrum of freely falling clouds prepared selectively in one of the three dressed states of the $F=1$ manifold (Section \ref{sec:exp-data-Nottingham}). In a second experiment
(Section \ref{sec:exp-data-Crete}), the spectrum is obtained for $^{87}$Rb atoms in the dressed $|1,-1\rangle$ state trapped in an optical potential, with particular focus on the group of transitions corresponding to $n=1$, as defined in Eq.\,(\ref{eq:ResonantCondition}). The third experimental configuration studies the MW spectrum of atoms confined in an RF-dressed shell trap (Section \ref{sec:atomsinashell}), where effects of the inhomogeneity of the field distribution play an important role. The experimental details for the different dressing configurations are presented in the following sections \ref{sec:exp-data-Nottingham}--\ref{sec:atomsinashell}, including analysis and discussion of the observed spectroscopic measurements.

\subsection{Free-falling atoms in homogeneous fields}
\label{sec:exp-data-Nottingham}

\begin{figure}[tb]
\centering
\includegraphics[width=\columnwidth]{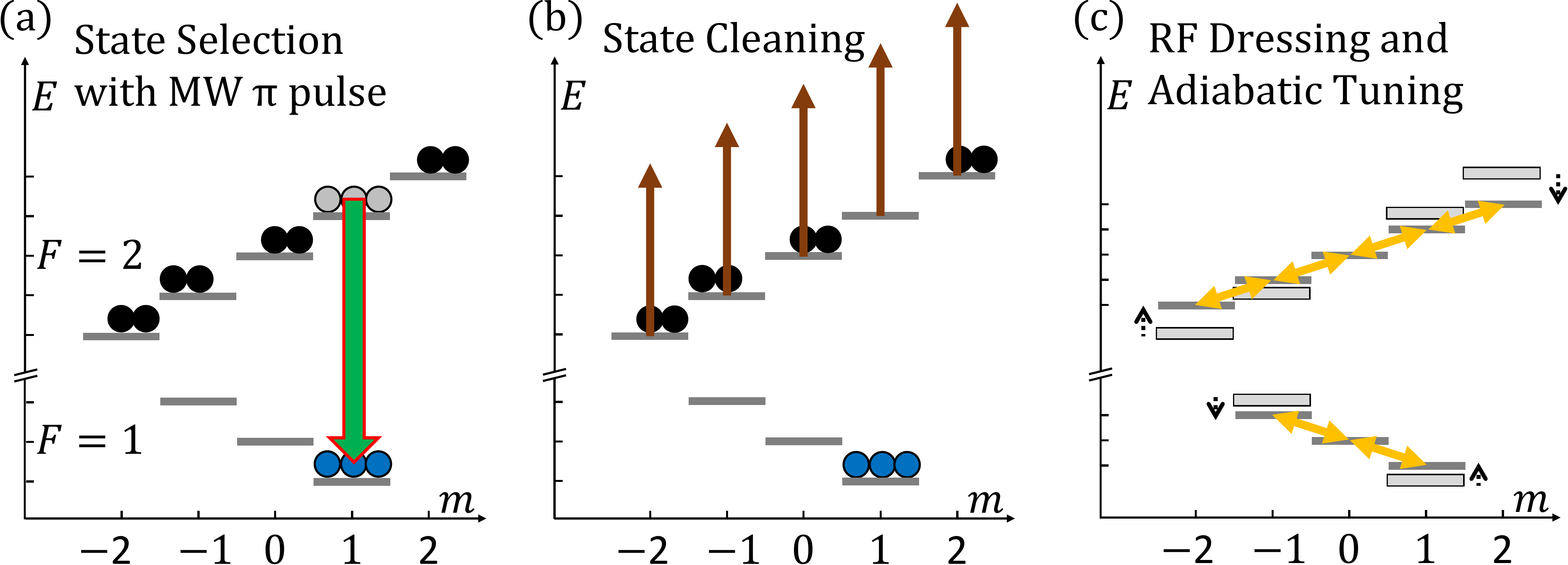}
\caption{\label{fig:StatePrep}Preparation sequence for a pure dressed state  $|F=1,\bar{m}\rangle$, with $\bar{m}=1$ for the specific example shown.
After an initial optical pumping stage, we start with atom population in $F=2$. We apply a MW $\pi$-pulse from $|2,m\rangle \rightarrow |1,m\rangle$ to selectively populate only one of the $F=1$ levels (a), before removing all atoms from the $F=2$ manifold with a resonant laser beam (b). Finally, we apply the dressing RF field and adiabatically tune the Larmor precession frequency into resonance by ramping the static field (c).} 
\end{figure}

Using free-falling ensembles of $^{87}$Rb atoms released from a magneto-optical trap (MOT) allows us to apply nearly homogeneous magnetic fields 
to otherwise unaffected atoms. By preparing pure dressed states and using a dispersive detection method to obtain state-dependent signals \cite{jammi_pra_2018} we are able to attribute spectroscopic features to individual transitions. 
The state preparation sequence, shown in Fig.~\ref{fig:StatePrep}, is performed after optical molasses cooling and optical hyperfine pumping with initial atomic population in all five Zeeman sub-levels of $F=2$.
We apply a MW $\pi$-pulse in a weak, homogeneous magnetic field ($\approx 1~\mathrm{G}$) by driving coherent Rabi cycles on one of the bare $\pi$-transitions. These transitions are non-degenerate due to the opposite sign of the $g$-factors in the two hyperfine states and thus frequency selective. This allows us to populate a single Zeeman sub-level in the $F=1$ manifold, i.e.\ only one of the states $|F=1,m=\pm 1,0\rangle$, see the example in Fig.~\ref{fig:StatePrep}a.
For each of the three $\pi$-transitions,
we adjust the pulse duration to maximise population in the target state. Subsequently, the population in the $F=2$ manifold is removed by shining a resonant laser beam tuned to the $F=2\rightarrow F'=3$ transition of the $D_2$-line (Fig.~\ref{fig:StatePrep}b). Multiple photon scattering on this closed transition
accelerates atoms away from the observed volume. Finally, the remaining atoms in the pure bare state are adiabatically dressed by ramping up the RF-field amplitude  and tuning the atomic Larmor frequency near resonance using the static field amplitude, see Fig.~\ref{fig:StatePrep}c.
For this set of experiments, we work in the weak field regime using a dressing field amplitude of $B_{\mathrm{RF}}\approx 10~\mathrm{mG}$ at a frequency of $\omega_{\textrm{RF}}=2\pi\times 180~\mathrm{kHz}$,
resonant for a static field of $B_\mathrm{DC}\approx 257~\mathrm{mG}$. The final dressed state is typically populated by $n_m\approx 3\times 10^7$ atoms.

The spectroscopy is performed by first applying a weak MW pulse, typically a few ms long, which may couple the prepared initial
dressed state in the $F=1$ manifold to one of the five
dressed states in the $F=2$ manifold, depending on the frequency of the MW pulse. 
The atomic response is then recorded by observing the AC-modulated linear birefringence of the ensemble, which we can measure separately for both hyperfine states using two laser beams and a balanced polarimeter \cite{jammi_pra_2018}.
An ensemble of atoms in a (bare) Zeeman state $|F,m\rangle$ will exhibit a linear birefringence $S$ proportional to atom number $n_m$. The birefringence depends quadratically on the magnetic quantum number $m$ and may change sign according to $S \propto n_m(F(F+1)-3m^2)$. 
Adiabatic dressing of the atoms modulates the linear birefringence of the ensemble, and depending on laser detuning and experimental geometry, we can detect a signal 
\begin{equation}
S_2\propto n_{\bar{m}}(F(F+1)-3\bar{m}^2)
\label{eq:statebirefringence}
\end{equation}
at the second harmonic of the dressing frequency, where sign and amplitude now depend on the adiabatic quantum number $\bar{m}$.

\begin{figure}[!htb]
\centering
\includegraphics[width=\columnwidth]{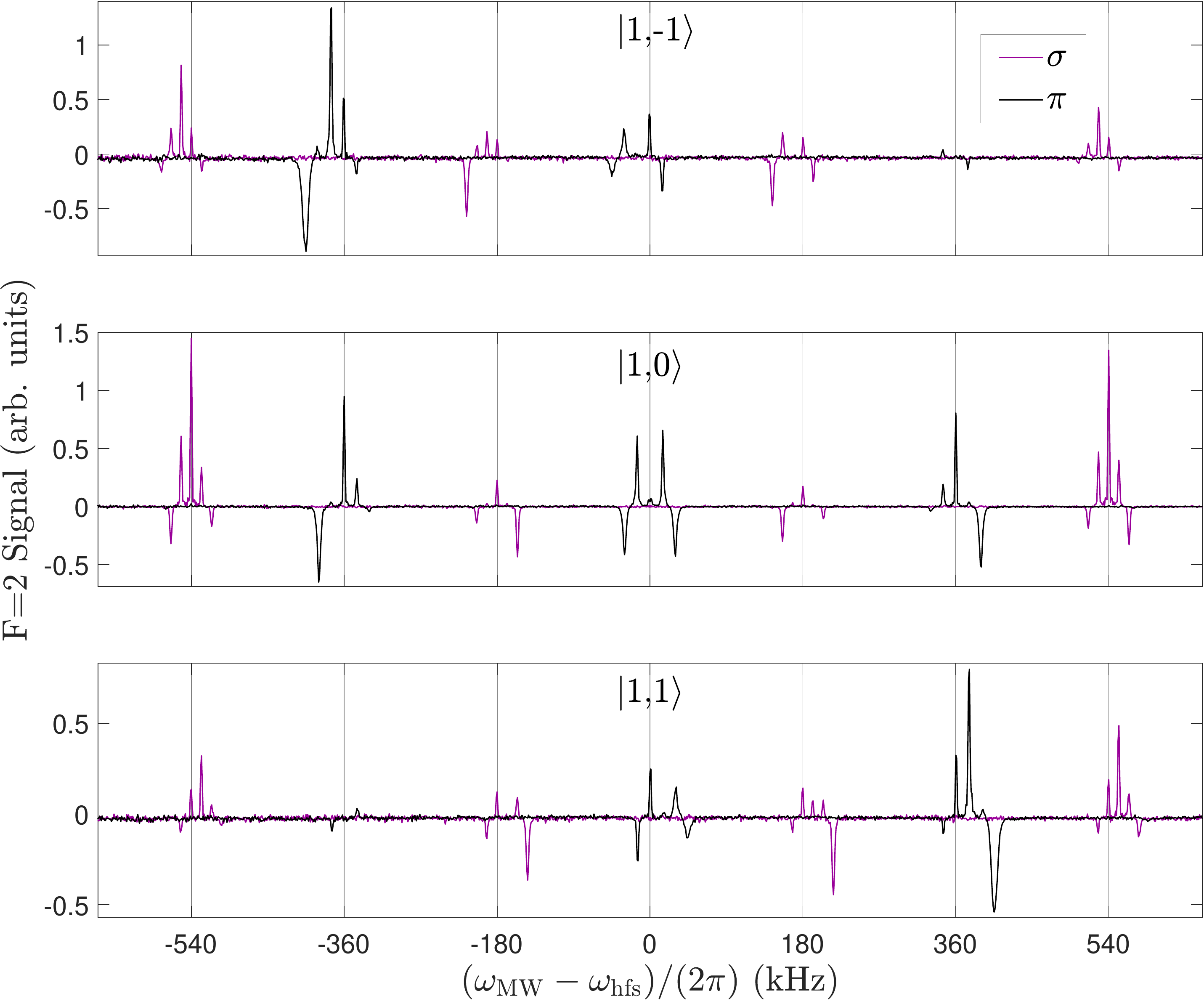}
\caption{\label{fig:full10spectrum}Experimental MW spectrum for RF dressed $^{87}$Rb showing the seven main spectral groups, corresponding to the bare hyperfine transitions. Each panel shows results for atoms prepared in one of the pure, initially bare states $|F=1,\bar{m}=\pm 1,0\rangle$, adiabatically dressed by a near resonant RF field with Rabi frequency $\Omega_\textrm{RF}\approx 2\pi\times10~\mathrm{kHz}$. The spectrum shows all transitions to $F=2$ for linear MW polarization both parallel ($\pi$) and orthogonal ($\sigma$) to the static field, with a resolution of 1 kHz. The $\sigma$ data set is taken with all fields (DC, RF, MW) pairwise orthogonal. The groups are separated by $\omega_{\textrm{RF}}=2\pi\times 180$~kHz, each showing transitions to the five dressed states of the $F=2$ manifold, where the outer transitions to dressed states $|2,\bar{m}=\pm2\rangle$ can be identified by negative signals. This data set confirms level assignments and expected frequencies as given by Eq.~(\ref{eq:ResonantCondition}), compare to Fig.~\ref{fig:TypicalSpectra87Rb}.  
}
\end{figure}

Depending on the polarization of the MW field, we observe up to seven main groups of dressed hyperfine transitions. As can be seen in Fig.~\ref{fig:full10spectrum}, each group is centred around one of the bare hyperfine transition frequencies, which are separated by the dressing frequency of
$\omega_{\textrm{RF}}= 2\pi\times180~$kHz. 
The appearance of the groups depends on the polarization of the MW field and resembles the bare scenario, with three groups emerging for $\pi$-polarization (i.e.\ $B_\textrm{MW}$ aligned with the static field $B_\textrm{DC}$), and four groups for linear $\sigma$-polarization (i.e.\ $B_\textrm{MW}$ orthogonal to $B_\textrm{DC}$). 
The frequencies of individual transitions are in good agreement with the theoretical prediction from Eq.~(\ref{eq:ResonantCondition}).
The individual peak heights and widths of the experimental data are not a direct reflection of the transitions' coupling strengths due to their dependence on various experimental settings. The widths of these lines are determined by a combination of MW power broadening, residual field inhomogeneities and magnetic field noise.
The experimental data shows some transitions that are predicted to vanish according to the approximation $g_1=-g_2$ that was used to produce the theoretical spectrum shown in Fig.~\ref{fig:TypicalSpectra87Rb}. These transitions are observable because the small difference in the magnitude of the Land\'e factors $g_1$ and $g_2$ and detuning(s) from RF resonances lead to non-zero coupling coefficients, see Eq.~(\ref{eq:MWCouplingofRFdressedStates}).

In our experiment, the population signals from the different $F=2$ levels scale relative to each other by a factor given by Eq.~(\ref{eq:statebirefringence}). 
The peak heights are not directly indicative of the transition strengths as the MW pulse of fixed duration (0.4 ms) induces Rabi cycles of differing frequencies for each transition and results in a different population fraction in $F=2$ depending on the number of Rabi cycles on each transition. 
The data for the three initial states differ in strength due to variations in the experimental state preparation efficiency, and the $\pi$- and $\sigma$- polarization data sets may be subject to variations in external experimental conditions as these were taken at different times.

\begin{figure}[!htb]
\centering
\includegraphics[width=\columnwidth]{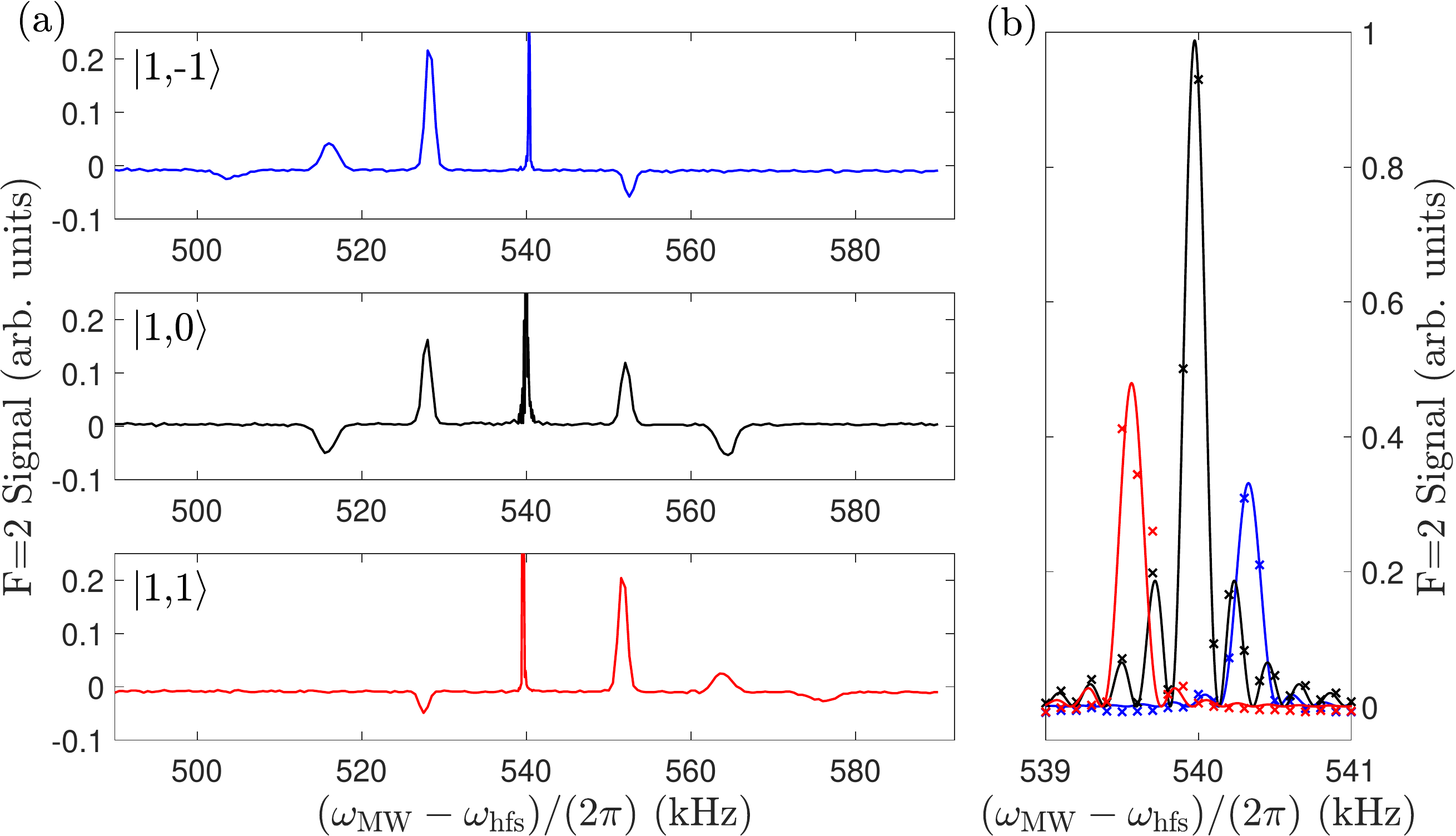}
\caption{\label{fig:1stgroup}Experimental $\sigma$-polarized MW spectra of RF dressed $^{87}$Rb atoms showing the group of $15$ transitions around $3 \times \omegaRF=2\pi\times540$~kHz detuning. The magnetic MW, RF and static fields are pairwise orthogonal. The panels in (a) show spectra for atoms are prepared in different sub-levels $|F=1,m=-1, 0,1\rangle$, adiabatically dressed under the same conditions as in Fig.~\ref{fig:full10spectrum} before shining a low power MW spectroscopy pulse of 5 ms duration and detecting atomic population in $F=2$. 
The slight negative offset from zero signal for states $m = \pm1$ is due to imperfect state preparation. The three sharpest peaks from (a) have a resolution of 0.1~kHz and are shown in (b) where blue, black and red crosses correspond to transitions from $|F=1, m=-1,0,1\rangle$ respectively. The solid lines model the data assuming only coherent driving. The small frequency shifts between the central peaks result from unequal magnitudes of the two $g_F$ factors. For more details on these peaks, see the main text.
}

\end{figure}

The set of transitions corresponding to the group of resonances in the vicinity of $\omega_{\textrm{hfs}}+3\times \omega_{\textrm{RF}}$ is shown in Fig.~\ref{fig:1stgroup}a. 
As before, atoms prepared in each of the initial three states give rise to
five resonant transitions separated in frequency by the RF Rabi frequency ($\approx 10~$kHz).
The strength of the signal reflects not only the MW transition strength, but also carries a signature of the populated target state in $F=2$ according to Eq.\,(\ref{eq:statebirefringence}), which explains why signals from transitions to $|F=2,m=\pm 2\rangle$ are negative in sign.
This spectrum was acquired
with low MW power in order to significantly reduce the effect of power broadening on the transition peaks. The width of the spectral lines in this case
is a consequence of homogeneous broadening due to field noise and inhomogeneous broadening due to magnetic field gradients for all but the three sharpest peaks. 
The sharp resonances in each group, shown in Fig.~\ref{fig:1stgroup}b,
correspond to the transitions
$|1,-1\rangle \rightarrow |2,1\rangle$, $|1,0\rangle \rightarrow |2,0\rangle$ and $|1,1\rangle \rightarrow |2,-1\rangle$. 
These transitions are least affected by the fields, because states in each pair experience (almost) equal magnetic shifts due to near identical factors $g_Fm_F$ for the involved states. A small frequency splitting between these transitions remains due to the marginal difference in magnitude of the $g_F$ factors. As a result, these lines are coherently driven, with theoretical line shapes of the form
\begin{equation}
A \propto \frac{\Omega^2}{\Omega^2+(\Delta-\Delta_c)^2}\sin^2{\frac{\sqrt{\Omega^2+(\Delta-\Delta_c)^2}t}{2}}
\label{eq:multiplerabi}
\end{equation}
where $A$ is the $F=2$ signal amplitude, $\Omega$ is the MW Rabi frequency, $\Delta_c$ is the centre frequency, $t$ is the pulse duration of 5~ms and $\Delta = \omega_\textrm{MW}-\omega_\text{hfs}$ \cite{foot_book_2004}.

In principle, the Rabi frequencies extracted from the least squares fit using Eq.~(\ref{eq:multiplerabi}) should allow for a comparison with the theory. Under the approximation $g_1=-g_2$, the theoretically predicted ratio of resonant coupling strengths is 1 : $-\sqrt{3}$ : 1 for the pairs with $\bar{m}=\mp1, 0, \pm1$, respectively, see outer groups in Fig.~\ref{fig:TypicalSpectra87Rb}. These ratios are qualitatively reflected by the experimental data. However, experimental uncertainties in the relative populations of the initial states as well as in the signal scale prohibit an accurate determination from just the line shapes.

\subsection{$^{87}$Rb in an optical dipole trap}
\label{sec:exp-data-Crete}

In the second set of experiments we confine the atoms in a crossed-beam optical dipole trap,
which allows us to work at high field strengths and address all dressed states in a trapped scenario.
The preparation sequence begins by loading an atom cloud from a  MOT into a magnetic quadrupole trap, where it is compressed and evaporatively cooled.
This is followed by further compression and evaporation in the crossed-beam dipole trap ($\lambda=1064$\,nm, P=1.8~W , final axial and radial trapping frequencies $\omega_\text{z}/2\pi \approx 180$\,Hz and $\omega_{\rho}/2\pi \approx 30$\,Hz).
This yields a fully polarized sample of approximately
$3 \times 10^5$ atoms at 50\,nK in the bare state $|1,-1\rangle$. At this stage, a vertical bias field $B_{\textrm{DC}}\ez$ is ramped from zero up to $|g_{F}| \mu_\text{B}B_{\textrm{DC}}\approx 1.4\,\hbar \omega_{\textrm{RF}}$, where $\omega_{\textrm{RF}}$ is the RF frequency of the dressing field that will be applied. We then switch on an RF-dressing field of frequency $\omega_{\textrm{RF}}$, which is linearly polarised along $\ex$, and subsequently dress the cloud by adiabatically ramping down $B_{\textrm{DC}}$ until a near-resonant condition $|g_{F}| \mu_\text{B} B_{\textrm{DC}}\approx\hbar \omega_{\textrm{RF}}$ is reached in $\Delta t=200~$ms. The spectroscopy is performed by shining a  microwave pulse of duration $\Delta t_{\textrm{MW}}=0.7$\,ms,
followed by a short free-expansion of typically 5~ms, right after all AC fields are switched off. This is followed by absorption imaging adapted for simultaneous recording of the atoms transferred to the $F=2$ manifold and atoms remaining in the $F=1$ manifold. 

\begin{figure}
\centering
\includegraphics[width=0.6\columnwidth]{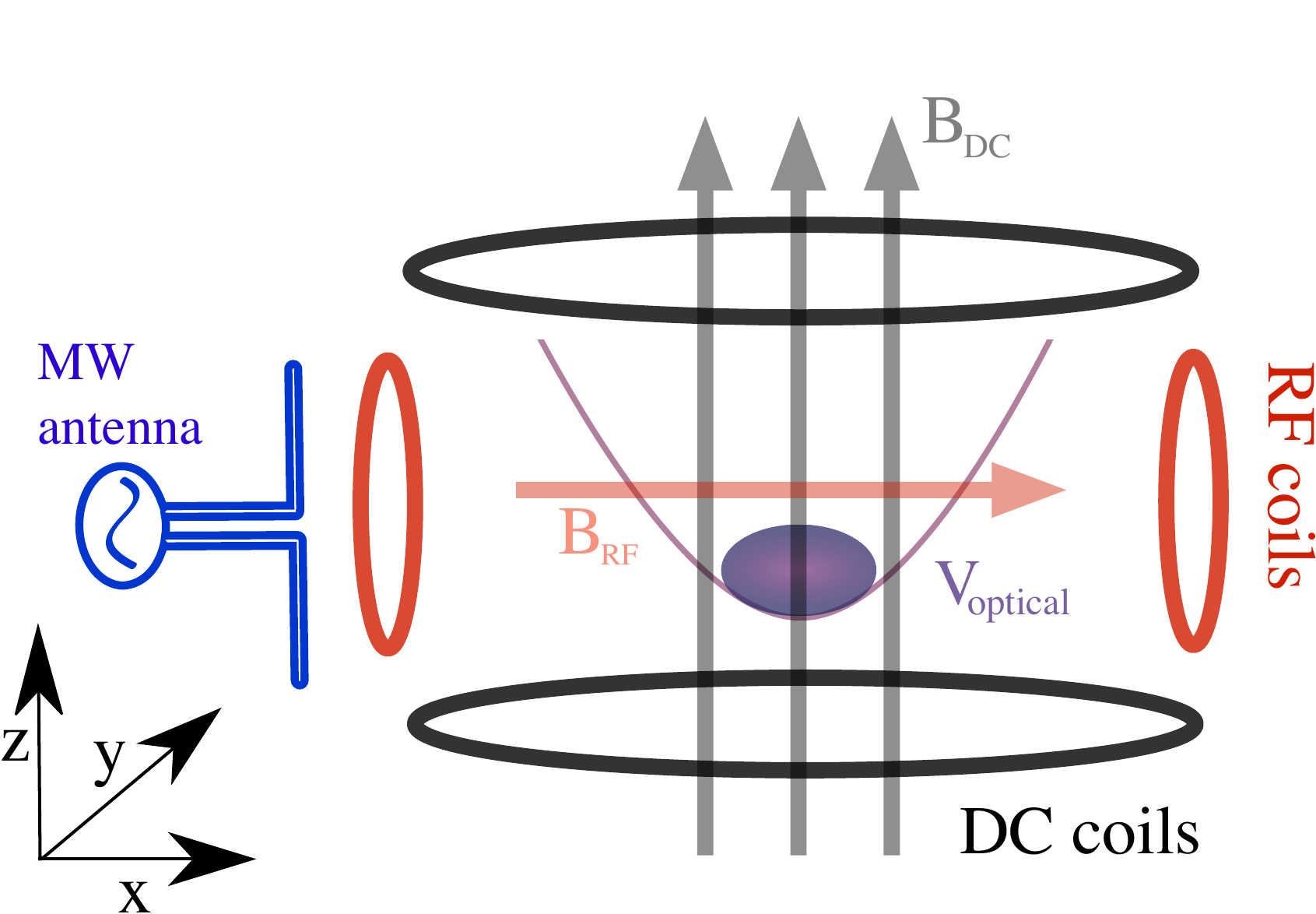}
\caption{\label{fig:opticaldipoleSetup}Schematic of the experimental setup with atoms in an optical dipole trap. $B_{\textrm{RF}}$ is generated from a pair of Helmholtz coils (in red, and labelled RF coils in the drawing) tuned with a resonant circuit, and points along the $x$-direction. $B_{\textrm{DC}}$ is generated from another set of coils (in black, and labelled ``DC coils'') that point along the $z$-direction. A MW dipole antenna produces a field approximately polarised in the $x$-$y$ plane. See the text for further details.}
\end{figure}

\begin{figure}
\centering
\includegraphics[width=\columnwidth]{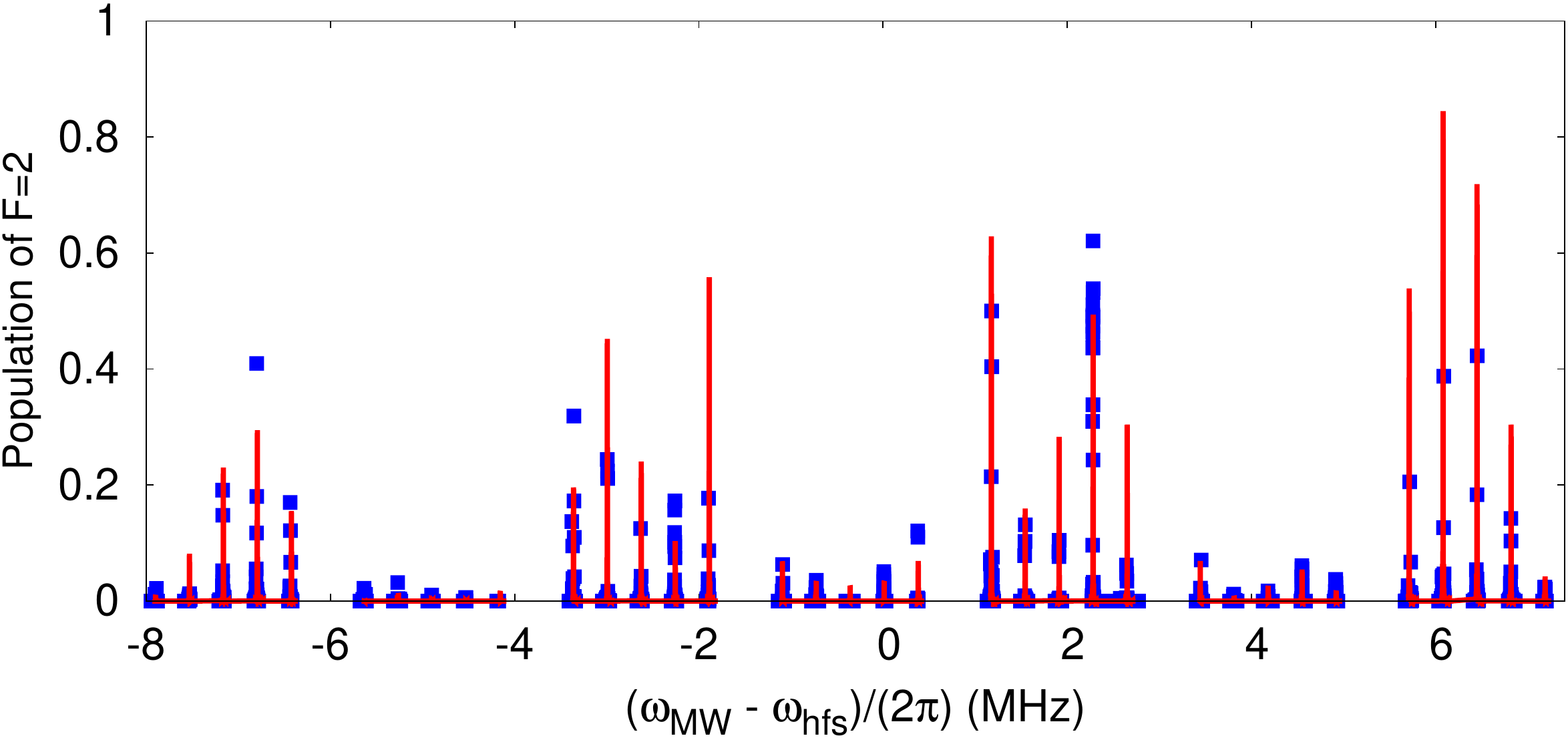}
\caption{\label{fig:cretespectrum}Full MW spectrum of RF-dressed $^{87}$Rb in an optical dipole trap. The dots correspond to experimental data and the lines shown numerical calculations. The initial sample is prepared in the dressed $|1,-1\rangle$. See text for details.}
\end{figure}

The RF-fields are produced by a pair of Helmholtz coils such that the generated  magnetic field points along $\ex$. We generate the MW field with a tuned dipole antenna placed in the $x$-$y$ plane, 
forming an angle of $45^\circ$ with the $\ex$ axis as we sketch in Fig.\,\ref{fig:opticaldipoleSetup}. 
The antenna was aligned to produce a MW-field linearly polarized in the $x$-$y$ plane, at $45^{\circ}$ from the $x$-axis and orthogonal to $B_{DC}$. 
The finite amplitude of the even groups in the MW spectroscopy results (Fig.\,\ref{fig:cretespectrum}) suggest that the MW-field polarization is not exactly orthogonal to $B_\text{DC}$ because of reflections from neighbouring metallic surfaces. The duration of the MW radiation pulse $\Delta t_{\textrm{MW}}$ was chosen to be much shorter than a $\pi/2$ pulse for the strongest transition. This allows direct comparison with the theoretical predictions for weak MW-fields from Section~\ref{sec:mw-coupling-rf}.

As in the case of the free-falling atoms (Section~\ref{sec:exp-data-Nottingham}), when the atoms are dressed and trapped in a crossed dipole potential, we observe seven groups of five transitions (for the initial state $|1,-1\rangle$) with variable couplings that depend on the configuration of the magnetic fields.
Fig.\,\ref{fig:cretespectrum} shows the full measured spectrum starting with a cloud prepared in the dressed state $|1,-1\rangle$ together with the theoretical prediction from Eqs.\,(\ref{eq:MWCouplingofRFdressedStates}) and (\ref{eq:ResonantCondition}).
The measured spectrum is for the field configuration described above. In this case, the MW antenna is oriented such that it produces a MW field that lies in the plane of the RF-field, mostly orthogonal to the static magnetic field. As a result of this MW-polarization, when we scan the MW frequency, the number of atoms transferred to the upper hyperfine manifold for the even groups is significantly smaller compared to the number of atoms transferred for the odd groups.

The vertical scale of Fig.\,\ref{fig:cretespectrum} shows the fraction of atoms transferred to the upper states starting from $F=1$. This is calculated from a separate measurement of the total atom number in the sample,
with $\omega_{\textrm{RF}}/2\pi=2.27$\,MHz. 
Quantitative agreement between the experimental results and the theoretical values is limited by other experimental factors not considered in this analysis: e.g.\ atomic losses, and drifts in the RF amplitude or in the homogeneous magnetic fields. 
Nevertheless, there is a good agreement between the theoretical predictions of the transition frequencies in Eqs.\,(\ref{eq:MWCouplingofRFdressedStates}) and (\ref{eq:ResonantCondition}) with our experimental results.
In particular, the peaks corresponding to the $\pi$-polarised component of the MW field are well reproduced by our theory, with qualitative agreement for the the circularly polarized components.

\begin{figure*}[t]
\centering
\includegraphics[width=0.7\textwidth]{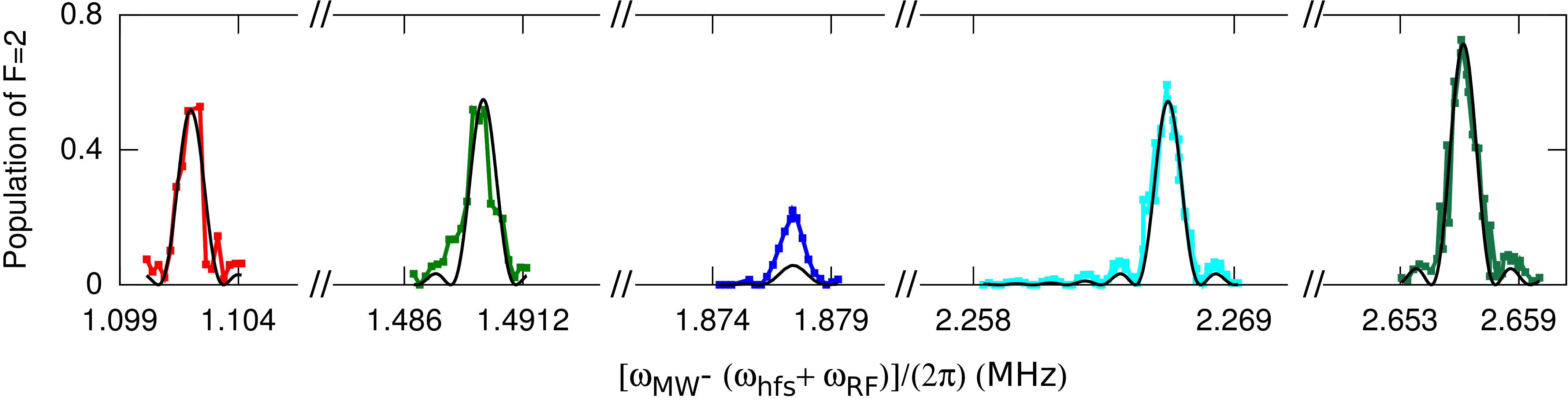}
\caption{\label{fig:cretespectrumGroup} MW spectra for the $+\omega_{\textrm{RF}}$ group of MW transitions with RF-dressed $^{87}$Rb. The initial state is the dressed $|1,-1\rangle$. Black solid curves represent a numerical fit of the transition probability to the data displayed in several solid colour curves, showcasing transitions spaced by $\Omega_{\textrm{RF}}$. In particular, they correspond to the following dressed states in $F=2: |2,-2\rangle$ (red), $|2,-1\rangle$ (green), $|2,0\rangle$ (blue), $|2,1\rangle$ (magenta) and $|2,2\rangle$ (turquoise). See text and Table~\ref{table:fieldfits} for details.}
\end{figure*}

These findings motivate the use of MW spectroscopy as a tool to determine the field configuration driving the atomic cloud. In order to test this idea, we took a spectrum of the group of resonances in the vicinity of $\omega_{\text{hfs}}+\omega_{\textrm{RF}}$ using $\omega_{\textrm{RF}}/2\pi=2.26341$\,MHz and  $\Delta t_{\textrm{MW}}=0.7$\,ms. 
Scanning the microwave frequency, we directly determine the transition probability by measuring the population of both hyperfine manifolds after the MW pulse. Calculating the numerically exact atomic time-evolution \cite{Sinuco2019}, we adjust the components of all applied fields to get the best fit to the experimental results.  We also adjust all three components of the static field since the Earth's magnetic field adds components in the $x$-$y$ plane in our set-up. We fit the $x$ and $y$ components of the microwave field because they produce significant couplings in the range of frequencies tested. The RF antennas are oriented to produce a RF field linearly polarised in the $x$ direction. The Table~\ref{table:fieldfits} shows the value of the parameters adjusted and Fig.~\ref{fig:cretespectrumGroup} shows a comparison of the experimental data with the fit. This procedure yields a measurement of the MW field amplitude $B_{\textrm{MW,x}}$ with a precision of approximately $10^{-2}$ and the error on $B_{\textrm{RF,x}}$ is of order $10^{-3}$. These errors depend on the knowledge of the DC-magnetic fields and the precision of the transition frequency measurement, which becomes worse for broader and more noisy line-shapes.

\begin{table}[!htb]
\begin{center}
\begin{tabular}{c|c}
Field component & Fitted value \\
\hline
\hline

$B_{\text{MW},x}$ &  $2.01 \pm 0.07~$mG \\
$B_{\text{MW},y}$ &  $1.33 \pm 0.08~$mG \\
$B_{\text{RF},x}$ &  $1.114 \pm 0.001~$G \\
$B_{\text{DC},x}$ &  $0.21 \pm 0.02~$G \\
$B_{\text{DC},y}$ &  $0.25 \pm 0.02~$G \\
$B_{\text{DC},z}$ &  $3.162 \pm 0.008~$G\\

\hline
\end {tabular}
\caption[]{\label{table:fieldfits}Values of the components of AC and DC magnetic fields  obtained from a a fit to the data in Fig.~\ref{fig:cretespectrumGroup}
}
\end{center}
\end{table}

\subsection{$^{87}$Rb in an  RF-dressed shell trap}
\label{sec:atomsinashell}

\begin{figure}[!hb]
\centering
\includegraphics[width=0.6\columnwidth]{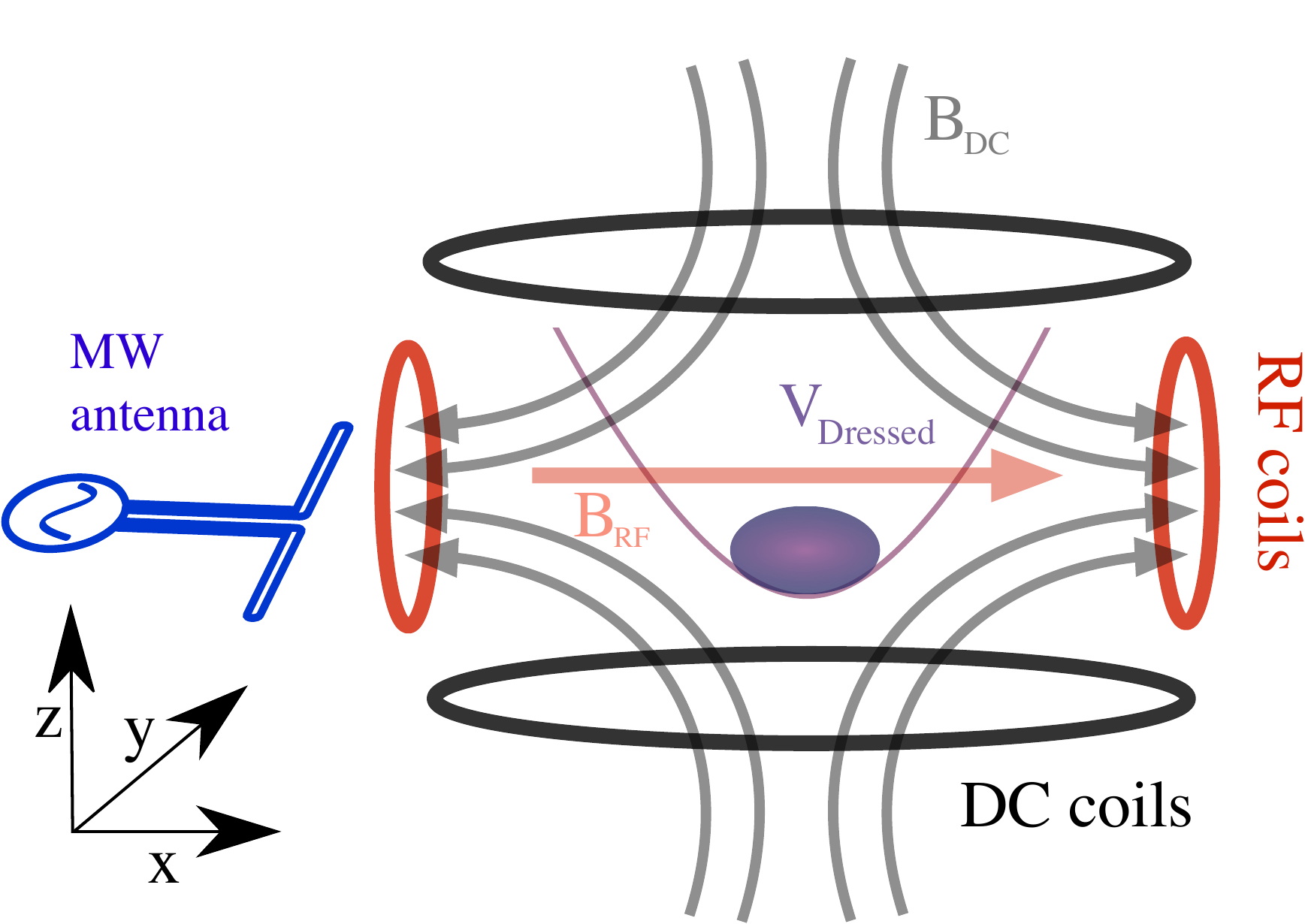}
\caption{\label{fig:RFshellExp}Schematic of the experimental setup with atoms in an RF-shell. The RF-field is generated as in Fig.~\ref{fig:opticaldipoleSetup}. The MW-field set-up is also the same, with a different tilt of the antenna. Atoms are trapped by a quadrupole magnetic field, instead of the optical field from Sec.~\ref{sec:exp-data-Crete}. This field is generated by a pair of anti-Helmholtz coils (in black, and labelled ``DC coils'') that are aligned in the $z$-direction. }
\end{figure}

We produce an RF-dressed shell trap
\cite{garraway_jphysb_2016,merloti_njp_2013} by modifying the current
in the DC coils of Fig.~\ref{fig:opticaldipoleSetup} so that it is now
in an anti-Helmholtz configuration as in Fig.~\ref{fig:RFshellExp}.
We apply an RF field as before.
When such an RF-dressed shell trap and a dipole trap are spatially matched through the resonant condition $|g_{F}| \mu_\text{B} B_{\textrm{DC}}=\hbar \omega_{\textrm{rf}}$, then the atom cloud in the dipole trap can be transferred to the shell trap by ramping up a quadrupole magnetic gradient and slowly ($\Delta t=0.5$\,s) ramping down to zero the power of the dipole beams (see Fig.~\ref{fig:RFshellExp}).
With this method, atoms can be loaded in the dressed $|1,-1\rangle$ state adiabatically, with non-measurable atom loss or heating. The shell trap potential can be written as \cite{zobay_prl_2001}
\begin{equation}
V_{|F,\bar{m}\rangle}\left(\mathbf{r}\right)=s\left(I+\frac{1}{2}\right)\frac{\hbar \omega_\text{hfs}}{2}+ s\bar{m} \hbar  \sqrt{\delta_{F}^{2} + \Omega_{\textrm{RF}}^{2}\left(\mathbf{r}\right)}+M g z ,
\label{eq:dressedPotentiala}
\end{equation}
with $g$ the gravitational acceleration, $M$ the atomic mass of $^{87}$Rb,  and the detuning $\delta_{F}=\Omega_\text{L}^{F}\left(\mathbf{r}\right)-\omega_{F}$ (with $|\hbar \Omega_\text{L}^{F}(\mathbf{r})| = |g_{F}| \mu_{B} |B_{\text{DC}}(\mathbf{r})|$, where $\Omega_\text{L}(\mathbf{r})$ is the Larmor frequency) and  $\Omega_{\textrm{RF}}(\mathbf{r})$ is the spatially dependent Rabi coupling \cite{perrin_advamop_2017}.
The parameter $s$ in Eq.~(\ref{eq:dressedPotentiala}) is given by $s=g_{F}/|g_{F}|$ so that $s=1$ for $F=2$ and $s=-1$ for $F=1$.

\begin{figure}[ht!]
\centering
\includegraphics[width=0.33\textwidth]{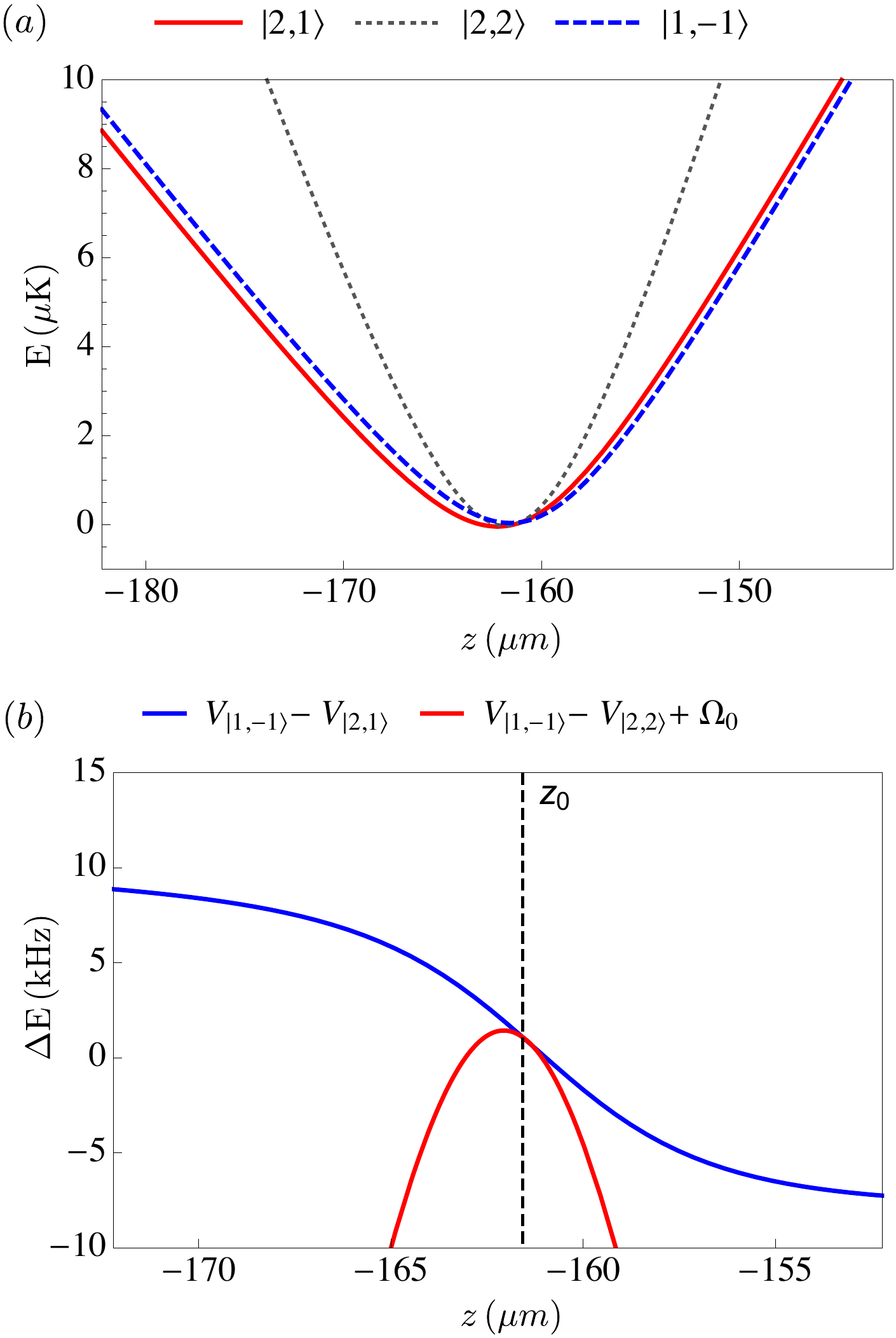}
\caption{\label{fig:twotrappabletransitions}(a) Dressed potentials $V_{|1,-1\rangle}$ (blue), $V_{|2,1\rangle}$ (red) and $V_{|2,2\rangle}-\Omega_{0}$ (dashed black) as calculated from Eq.\,(\ref{eq:dressedPotentiala}). (b) Energy differences $\Delta E_1=V_{|1,-1\rangle}-V_{|2,1\rangle}$ (blue) and $\Delta E_2=V_{|1,-1\rangle}-V_{|2,2\rangle}+\Omega_{0}$ (red). The black dashed line labelled $z_{0}$ indicates the trap position of the \emph{initial} state $|1,-1\rangle$. We consider a quadrupole gradient $\alpha=100$\,G/cm, radio-frequency $\omega_{\textrm{RF}}/2\pi=2.25891$\,MHz and a linearly polarised RF-field with $B_{\textrm{RF}}=0.2$\,G.}
\end{figure}

Trappable states are those where $g_{F} \bar{m}>0$. 
As we present in Fig.\,\ref{fig:twotrappabletransitions}, this leads to state dependent traps, not only with regards to the RF-polarization coupling $g_{F}$-dependence, but also on the quadrupole-field induced $\bar{m}$-dependent force.
Concretely, in Fig.\,\ref{fig:twotrappabletransitions}a we show the trapping potentials for the three trappable states $|1,-1\rangle$, $|2,1\rangle$, $|2,2\rangle$. 
One can readily see that the traps have different curvatures and minima. 
In addition, in Fig.\,\ref{fig:twotrappabletransitions} we show the differences in energy $\Delta E_1=V_{|1,-1\rangle}-V_{|2,1\rangle}$ and $\Delta E_2=V_{|1,-1\rangle}-V_{|2,2\rangle}+\Omega_{0}$, which serve as an illustration of the inhomogeneous broadening related to the mismatch of the traps that a cloud of size $\Delta z$ would experience if such transitions were driven (with $\Omega_{0}$ as the Rabi frequency at the centre of the shell trap). 
One observes that, at the trap position $z_{0}$ of $V_{|1,-1\rangle}$ (the initial state), the curve $\Delta E_1$ is sloped, which is a direct result of the different $g_{F}$ factors. 
One can also see how the parabola-shaped curve $\Delta E_2$ is, firstly, not centred at $z_0$ (this is, again, due to the different $g_{F}$ factor); and, secondly, shows a larger curvature as $|z|$ diverges from the trap centre (this is a result of the different $\bar{m}$). 

\begin{figure}
\centering
\includegraphics[width=\columnwidth]{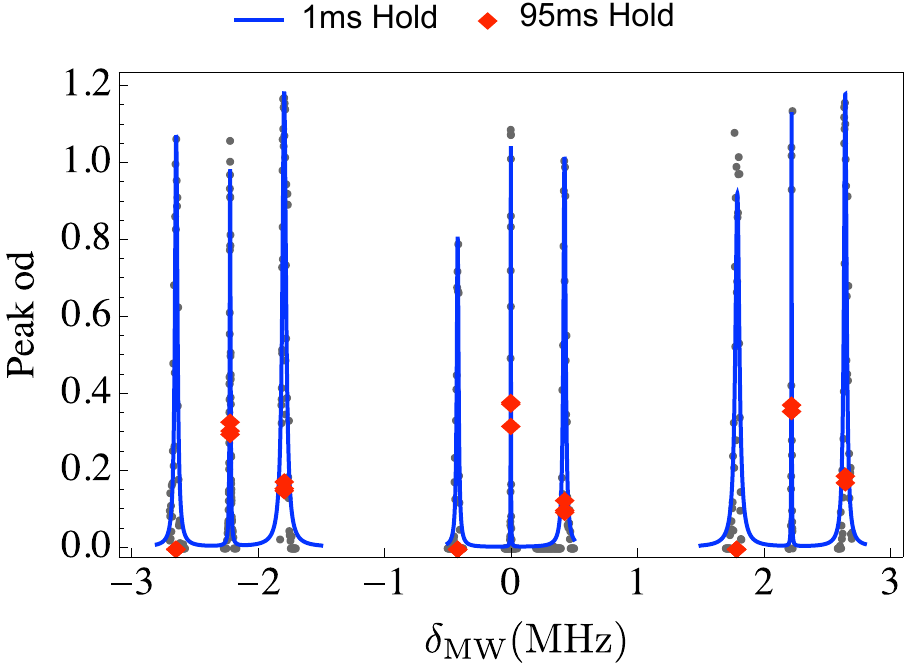}
\caption{\label{fig:shellLifetimes}The solid blue lines show Lorentzian curves fitted to the MW spectral measurements in F=2 (black dots) from the initial $|1,-1\rangle$ state. Red diamonds indicate the measured peak optical density at each of the resonant frequencies 95\,ms after the transfer. The horizontal axis shows the detuning of the microwave frequency $\delta_{\textrm{MW}}$ from the hyperfine frequency $\omega_{\textrm{hf}}/2 \pi$. The quadrupole gradient is $45$\,G/cm, the RF-frequency is $\omega_{\textrm{RF}}=2.22$\,MHz and the RF-field Rabi frequency is $\Omega_{\textrm{RF}}/2\pi=423\pm 2$\,kHz.}
\end{figure}

In the RF-dressed shell trap we observe the same MW spectrum structure found in Fig.\,\ref{fig:cretespectrum}. In this case, the trap geometry, its spatial location and the trapping frequencies are directly determined by the resonant condition of the RF-field and the DC magnetic quadrupole field, although the gravitational sag may become non-negligible. 
This results in state-dependent traps for any pair of initial and final states, which are in general different for different (trappable) states, as we showed in Fig.\,\ref{fig:twotrappabletransitions}. 
As a consequence, the transition line-width may increase in the magnetic trap (compared to the optical trap) and the transferred atoms will experience higher heating rates as they are coupled via MW radiation if the traps of the initial and final states lie in different positions. Moreover, any homogeneous magnetic DC field simply translates the quadrupole in space and thus does not shift transition frequencies. This is a consequence of the fact that the trap position is fundamentally determined by the resonant condition of the quadrupole field with the RF-dressing frequency: i.e.\ $|g_{F}|\mu_{B}\alpha z=\hbar \omega_{\textrm{RF}}$, where $\alpha$ is the quadrupole field gradient. In Fig.\,\ref{fig:shellLifetimes}, we show experimental measurements of the three central pairs of transitions ($\Delta t_{\textrm{MW}}=2.5$\,ms) from $|1,-1\rangle$ to $|2,0\rangle$, $|2,1\rangle$ and $|2,2\rangle$ for an adiabatic magnetic potential with Rabi frequency $\Omega_{\textrm{RF}}/2 \pi\approx 423\pm 2$\,kHz (in $\ex$), quadrupole gradient $\alpha = 45$\,G/cm and $\omega_{\textrm{RF}}/2 \pi=2.22$\,MHz. We have fitted simple Lorentzian curves to the spectral data after 1\,ms hold time (blue). We have furthermore measured the peak optical density after 95\,ms hold time for each of the transitions and we observe how the transitions from $|1,-1\rangle$ at $\omega_{\textrm{hfs}}+(n \omega_{\textrm{RF}}-\Omega_{\textrm{RF}})$ lead to the non-trapped state $|2,0\rangle$, at $\omega_{\textrm{hfs}}+n \omega_{\textrm{RF}}$ to $|2,1\rangle$ with 100\,ms lifetime and at $\omega_{\textrm{hf}}+(n \omega_{\textrm{RF}}+\Omega_{\textrm{RF}})$ to $|2,2\rangle$ with a 60\,ms lifetime. In both trapped states we observe significant heating due to the mismatch of the traps, being higher in the $|2,2\rangle$ case.

The line-widths are remarkably different for the three pairs of transitions because of the different overlap between the initial and final adiabatic potentials. In this experiment, the transition from $|1,-1\rangle$ to $|2,1\rangle$ is narrower ($1.5$\,kHz) than the other two transitions to $|2,0\rangle$ and $|2,2\rangle$, which are broader and more noisy ($20$\,kHz).

\section{Conclusions}
\label{sec:conclusions}

In this work we presented a complete theoretical and experimental study of the hyperfine spectrum of $^{87}$Rb dressed by an RF field. The theoretical analysis of the spectrum considers the regime of weak static and RF dressing fields. In all three experimental situations discussed, the overall features of the spectrum are well described by this analytic treatment. In particular, we found the relative position of the resonant frequencies and various selection rules associated with the polarization of the microwave probing field. In the case of free-falling atomic ensembles, the strengths of the applied fields are in the weak field regime and we identify all possible microwave transitions between pairs of radio-frequency dressed states. In this case, using the AC-modulated linear birefringence of the atomic ensemble prepared in fully polarised atomic states allows us to unambiguously assign quantum numbers and confirm the predicted value of the relative coupling strengths for all observed resonances. In the cases of atomic ensembles in the crossed-dipole and adiabatic shell traps, we used relatively strong DC ($\sim3.1$\,G) and RF fields ($\sim0.5$\,G). The number and distribution of allowed transitions remains the same as in our first experiment. However, the line-spacing is modified due to non-linear Zeeman shifts, which we include when fitting the measured spectrum. Finally, in the  case of the ensemble trapped in an adiabatic shell, the nature of the RF-dressed adiabatic potential leads to small spin-dependent discrepancies in the size and curvature of the trapping potential. Even though the spectrum remains unchanged, the lifetimes and heating rates in the shell trap depend strongly on the spin-states involved in the transition.

The study and experimental observation of the MW spectroscopy in RF-dressed states is a first step towards the characterisation and implementation of several quantum optics and atom interferometry schemes, such as the matter-wave interferometry in ring traps \cite{Pandey2019,stevenson_prl_2015} and atomic-clocks \cite{kazakov_pra_2015}. The experimental situations tested in this work have potential advantages for such applications. For example, trapped atomic ensembles permit interferometric sequences with long interrogation times, whereas collisions in free-falling ensembles can be exploited to increase the coherence time  using spin self-rephasing \cite{PhysRevLett.105.020401}. In all cases, it should be possible to find optimal dressing configurations that enable robust coherent manipulations between dressed states. Also, the sensitivity of the microwave spectrum to the polarization of the RF and MW fields can be used for precise measurements.  Finally, applications similar to those discussed here are currently being developed with a great variety of atomic and solid-state alkali-like systems (e.g.\ alkali-metals \cite{garraway_jphysb_2016}, alkali-metal-like ions \cite{PhysRevA.91.012322} and NV centres \cite{PhysRevLett.113.237601}), where similar spectral signatures can be observed and explained
using the theoretical framework we have presented.

The datasets generated for this paper are accessible at 10.17639/nott.7002 Nottingham Research Data Management Repository \cite{Datasets}.

\section*{Acknowledgements}
\label{sec:ack}

The UK authors thank the UK Engineering and Physical Sciences Research Council for support (Grant EP/M013294/1). 
The Greek part of this work is supported by the project \emph{HELLAS-CH} (MIS 5002735), which is implemented under the \emph{Action for Strengthening Research and Innovation Infrastructures}, funded by the Operational Programme \emph{Competitiveness, Entrepreneurship and Innovation} (NSRF 2014-2020) and co-financed by Greece and the European Union (European Regional Development Fund). 
GV received funding from the European Union's Horizon 2020 research and innovation programme under the Marie Sklodowska-Curie Grant Agreement No. 750017.
SP acknowledges financial support from the Hellenic Foundation for Research and Innovation (HFRI) and the General Secretariat and Technology (GSRT), under the HFRI PhD Fellowship grant (4823).

BG, WK and TF conceived  the main ideas and devised the project and WK initiated the collaboration.
HM, SP, BF, SJ and KP carried out the experiments.
GS, HM, SP, WK, BF, SJ, and KP contributed to the data analysis.
GS, BG, HM and TF are responsible for the theoretical work. 
HM, SP, VB, WK, BF, SJ, KP and TF contributed to building the experiments.
GS, BG, HM, SP, GV, WK, BF, SJ, KP and TF contributed to the result discussion and manuscript writing.

\appendix
\begin{widetext}
\section{\label{app:dmatrices}Matrix representation of the operator $\hat U_y=\exp(-i\theta \hat F_y)$ for total angular momentum F=1,2}

For $^{87}$Rb, the RF-dressed states described in Sec.\,\ref{sec:RWA} become linear superpositions of the Zeeman split states as in Eq.\,(\ref{eq:dressedbare}). The coefficients of such a superposition involve the Wigner $d$-matrix of
Eq.(\ref{eq:wignersmalldmatrix}). In the case of  $^{87}$Rb these are explicitly given by: 
\begin{equation}
d^{1}(\tdF)=
\begin{pmatrix}
\frac{\mathrm{cos}\left( \tdF\right) +1}{2} & −\frac{\mathrm{sin}\left( \tdF\right) }{\sqrt{2}} & \frac{1−\mathrm{cos}\left( \tdF\right) }{2} \cr 
\frac{\mathrm{sin}\left( \tdF\right) }{\sqrt{2}} & \mathrm{cos}\left( \tdF\right)  & −\frac{\mathrm{sin}\left( \tdF\right) }{\sqrt{2}} &\cr 
\frac{1−\mathrm{cos}\left( \tdF\right) }{2} & \frac{\mathrm{sin}\left( \tdF\right) }{\sqrt{2}} & \frac{\mathrm{cos}\left( \tdF\right) +1}{2} \cr
\end{pmatrix}
\label{eq:d1}
\end{equation}
with $F = 1$,  and 
\begin{equation}
d^{2}(\tuF)=
\begin{pmatrix}
\frac{{\left( \mathrm{cos}\left( \tuF \right) +1\right) }^{2}}{4} & \frac{\left( \mathrm{cos}\left( \tuF \right) +1\right) \,\mathrm{sin}\left( \tuF \right) }{2} & \frac{\sqrt{3}\,{\mathrm{sin}\left( \tuF \right) }^{2}}{{2}^{\frac{3}{2}}} & \frac{\left( 1−\mathrm{cos}\left( \tuF \right) \right) \,\mathrm{sin}\left( \tuF \right) }{2} & \frac{{\left( 1−\mathrm{cos}\left( \tuF \right) \right) }^{2}}{4}\cr 
−\frac{\left( \mathrm{cos}\left( \tuF \right) +1\right) \,\mathrm{sin}\left( \tuF \right) }{2} & \frac{2\,{\mathrm{cos}\left( \tuF \right) }^{2}+\mathrm{cos}\left( \tuF \right) −1}{2} & \frac{\sqrt{3}\,\mathrm{sin}\left( 2\,\tuF \right) }{{2}^{\frac{3}{2}}} & \frac{−2\,{\mathrm{cos}\left( \tuF \right) }^{2}+\mathrm{cos}\left( \tuF \right) +1}{2} & \frac{\left( 1−\mathrm{cos}\left( \tuF \right) \right) \,\mathrm{sin}\left( \tuF \right) }{2}\cr
 \frac{\sqrt{3}\,{\mathrm{sin}\left( \tuF \right) }^{2}}{{2}^{\frac{3}{2}}} & −\frac{\sqrt{3}\,\mathrm{sin}\left( 2\,\tuF \right) }{{2}^{\frac{3}{2}}} & \frac{3\,{\mathrm{cos}\left( \tuF \right) }^{2}−1}{2} & \frac{\sqrt{3}\,\mathrm{sin}\left( 2\,\tuF \right) }{{2}^{\frac{3}{2}}} & \frac{\sqrt{3}\,{\mathrm{sin}\left( \tuF \right) }^{2}}{{2}^{\frac{3}{2}}}\cr 
 −\frac{\left( 1−\mathrm{cos}\left( \tuF \right) \right) \,\mathrm{sin}\left( \tuF \right) }{2} & \frac{−2\,{\mathrm{cos}\left( \tuF \right) }^{2}+\mathrm{cos}\left( \tuF \right) +1}{2} & −\frac{\sqrt{3}\,\mathrm{sin}\left( 2\,\tuF \right) }{{2}^{\frac{3}{2}}} & \frac{2\,{\mathrm{cos}\left( \tuF \right) }^{2}+\mathrm{cos}\left( \tuF \right) −1}{2} & \frac{\left( \mathrm{cos}\left( \tuF \right) +1\right) \,\mathrm{sin}\left( \tuF \right) }{2}\cr 
 \frac{{\left( 1−\mathrm{cos}\left( \tuF \right) \right) }^{2}}{4} & −\frac{\left( 1−\mathrm{cos}\left( \tuF \right) \right) \,\mathrm{sin}\left( \tuF \right) }{2} & \frac{\sqrt{3}\,{\mathrm{sin}\left( \tuF \right) }^{2}}{{2}^{\frac{3}{2}}} & −\frac{\left( \mathrm{cos}\left( \tuF \right) +1\right) \,\mathrm{sin}\left( \tuF \right) }{2} & \frac{{\left( \mathrm{cos}\left( \tuF \right) +1\right) }^{2}}{4}
\end{pmatrix}
\label{eq:d2}
\end{equation}
with \[\theta_{F} = \frac{\pi}{2} - \tan^{-1} \left( \frac{ B_{\textrm{DC}} -  \hbar \omega_{\textrm{RF}}/(\mu_B |g_F|)}{\sqrt{2} B_{\textrm{RF},\textrm{sgn}{g_F}}} \right) . \]

These expressions simplify in case of resonant RF dressing where $\theta_F = \pi/2$.

\section{Matrix elements of the MW coupling in the basis of RF-dressed states}
\label{sec:matrix-elements-detail}

In the lab frame of reference, the polar decomposition of the MW coupling has the form
\begin{equation}
\hat{H}_{\textrm{MW}} = \sum_{\sigma\in{+,-,0}} \hat{H}_{\textrm{MW}}^\sigma  ,
\end{equation}
with
\begin{equation}
\hat{H}_{\textrm{MW}}^\sigma = \eta_\sigma \mu_\text{B} g_J \left(B_{\textrm{MW},\sigma}e^{-i \omega_{\textrm{MW}} t}\hat{J}_\sigma +  B_{\textrm{MW},\sigma}^*e^{i \omega_{\textrm{MW}} t} \hat{J}_{-\sigma}\right) ,
\end{equation}
where we used the definition $\eta_{+1}=-1/\sqrt{2}$, $\eta_{-1}=1/\sqrt{2}$ and $\eta_0 = 1$. Expressed as a sum of spherical angular momentum operators, these components of the Hamiltonian can be written as
\begin{equation}
\hat{H}_{\textrm{MW}}^\sigma = \mu_\text{B} g_J \eta_\sigma \sum_{\ell\in{+,-,0}} B_{\textrm{MW}}^{\ell,\sigma}(t)\hat{J}_\ell ,
\end{equation}
with
\begin{equation}
B_{\textrm{MW}}^{\ell,\sigma}(t) = \left(B_{\textrm{MW},\sigma}\left(\frac{1+\sigma\ell}{2}\right) + B_{\textrm{MW},\sigma}^*\left(\frac{1-\sigma\ell}{2}\right) \right) e^{-i\sigma\ell\omega_{\textrm{MW}}t} + (1-|\sigma|)\delta_{\ell,0}\left(B_{\textrm{MW},\sigma} e^{-i\omega_{\textrm{MW}}t} + B_{\textrm{MW},\sigma}^*  e^{i\omega_{\textrm{MW}}t} \right) .
\end{equation}
\begin{equation}
\hat{\bar{H}}_{\textrm{MW}}^\sigma = \frac{\mu_\text{B} g_J}{2} \eta_\sigma \sum_{\ell\in{+,-,0}} B_{\textrm{MW}}^{\ell,\sigma}(t)\hat{\bar{J}}_\ell ,
\label{eq:apx1}
\end{equation}
with
\begin{equation}
\hat{\bar{J}}_\ell =  \hat{U}_y^\dagger(\tu ,\td)\hat{U}_z^\dagger(\omega_{\textrm{RF}}t) \hat{J}_\ell \hat{U}_z(\omega_{\textrm{RF}}t) \hat{U}_y(\tu ,\td) .
\end{equation}

For concreteness, let's consider an element that couples states in different hyperfine manifolds:
\begin{align}
\left\langle \Fu,\bar{m}'\right|\hat{\bar{J}}_{\ell}\left|\Fd,\bar{m} \right\rangle = \sum_{m'=-\Fu}^{\Fu} \sum_{m =-\Fd}^{\Fd} &\left\langle \Fu,\bar{m}'\right| \hat{U}_y^\dagger(\tu )\hat{U}_z^\dagger(\omega_{\textrm{RF}}t) \left|\Fu,m'\right\rangle \nonumber \\
&\times  \left\langle \Fu, m' \right|  \hat{J}_{\ell} \left|\Fd ,m\right\rangle \left\langle \Fd, m \right| \hat{U}_z(\omega_{\textrm{RF}}t) \hat{U}_y(\td ) \left|\Fd,\bar{m} \right\rangle ,
\end{align}
in which we have used the identity operator of each hyperfine manifold in the lab frame $\hat{\mathds{1}}_F = \sum_m \left|F,m \right\rangle \left\langle F,m \right|$. Since the time-dependent rotation operator is diagonal in this basis we obtain
\begin{align}
\left\langle \Fu,\bar{m}'\right| \hat{\bar{J}}_\ell \left|\Fd,\bar{m} \right\rangle = \sum_{m'=-\Fu}^{\Fu} \sum_{m = -\Fd}^{\Fd} &\left\langle \Fu,\bar{m}'\right| \hat{U}_y^\dagger(\tu )\left|\Fu,m'\right\rangle e^{im' \omega_{\textrm{RF}}t} \nonumber \\
&\times \left\langle \Fu, m' \right|  \hat{J}_{\ell} \left|\Fd ,m\right\rangle e^{im \omega_{\textrm{RF}}t} \left\langle \Fd, m \right| \hat{U}_y(\td ) \left|\Fd,\bar{m} \right\rangle .
\end{align}
Now, using the the matrix representation of the rotation $\hat{U}_y$ given by the Wigner $d$-matrix \cite{messiah_book_2003} and rearranging the exponential factors we obtain:
\begin{equation}
\left\langle \Fu,\bar{m}'\right| \hat{\bar{J}}_\ell \left|\Fd,\bar{m} \right\rangle = \sum_{m'=-\Fu}^{\Fu} \sum_{m = -\Fd}^{\Fd} d^{\Fu}_{\bar{m}',m'} (-\tu ) e^{i(m' +m)\omega_{\textrm{RF}}t} d^{\Fd}_{m,\bar{m}} (\td )  \left\langle \Fu, m' \right|\hat{J}_{\ell} \left|\Fd ,m\right\rangle .
\end{equation}
Now we use the matrix elements of the electronic angular momentum operators, $J_\ell$, defined in terms of 3-j symbols \cite{messiah_book_2003} to obtain:
\begin{equation}
\left\langle \Fu,\bar{m}'\right| \hat{\bar{J}}_\ell \left|\Fd,\bar{m} \right\rangle \sqrt{\frac{2I(I+1)}{2I+1}}  \sum_{m'=-\Fu}^{\Fu} \sum_{m =- \Fd}^{\Fd} d^{\Fu}_{\bar{m}',m'} (-\tu ) e^{i(m' +m)\omega_{\textrm{RF}}t} d^{\Fd}_{m,\bar{m}} (\td ) (-1)^{(\Fu - m')} 
\begin{pmatrix}
\Fu & 1 & \Fd \\
-m'& \ell & m
\end{pmatrix}
\end{equation}
The 3j-symbols are different from zero if and  only if $-m'+\ell+m=0$, which help us to reduce one of the sums in the following way:
\begin{equation}  
\left\langle \Fu,\bar{m}'\right| \hat{\bar{J}}_\ell \left|\Fd,\bar{m} \right\rangle =\sqrt{\frac{2I(I+1)}{2I+1}}  \sum_{m = - \Fd}^{\Fd} d^{\Fu}_{\bar{m}',m+\ell} (-\tu ) e^{i(2m +\ell)\omega_{\textrm{RF}}t} d^{\Fd}_{m,\bar{m}} (\td ) (-1)^{(\Fu - m - \ell)} 
\begin{pmatrix}
\Fu & 1 & \Fd \\
-(m+\ell)& \ell & m
\end{pmatrix} .
\label{eq:AngularConservation}
\end{equation}
Putting this result together with Eq.~(\ref{eq:apx1}), we obtain
\begin{align}
\left\langle \Fu,\bar{m}'\right|\hat{\bar{H}}^\sigma_\text{MW}\left|\Fd,\bar{m} \right\rangle &= \eta_\sigma  \mu_\text{B} g_J \sqrt{\frac{2I(I+1)}{2I+1}} \sum_{\ell=-1}^1  B^{\sigma,\ell}_{\textrm{MW}}(t) \nonumber \\
&\times\sum_{m=-\Fd }^{\Fd } e^{i \omega_\textrm{RF}t(2m+\ell))} \times d^{\Fu}_{\bar{m}',m+\ell}(-\tu)d^{\Fd }_{m,\bar{m}}(\td) \nonumber \\
&\times (-1)^{(\Fu-m-\ell)}
\begin{pmatrix}
\Fu & 1 & \Fd \\
-(m + \ell)& \ell & m
\end{pmatrix} ,
\label{eq:MWCouplingofRFdressedStatesAPX}
\end{align}
as in Eq.~(\ref{eq:MWCouplingofRFdressedStates}).

We can also obtain explicit expressions for the couplings associated to each polar component of the microwave field oscillating at different frequencies ($\omega_{\textrm{MW}} + n \omega_\textrm{RF}$), following the factorisation of the coupling matrices in Eq.~(\ref{eq:H_MW_dec}):
\begin{eqnarray}
\left\langle \Fu,\bar{m}'\right| \hat{\bar{H}}^{-,-(2m-1)}_\textrm{MW} \left|\Fd,\bar{m} \right\rangle &=& \frac{\mu_\text{B} g_J B_{\textrm{MW},-}}{\sqrt{2}}   \sqrt{\frac{2I(I+1)}{2I+1}}  d^{\Fu}_{\bar{m}',m-1}(-\tu)d^{\Fd }_{m,\bar{m}}(\td) (-1)^{(\Fd+m)}
\begin{pmatrix}
\Fu & 1 & \Fd \\
-(m-1) & -1 & m
\end{pmatrix}, 
\nonumber\\
\left\langle \Fu,\bar{m}'\right| \hat{\bar{H}}^{+,-(2m+1)}_\textrm{MW} \left|\Fd,\bar{m} \right\rangle &=& -\frac{\mu_\text{B} g_J B_{\textrm{MW},+}}{\sqrt{2}}   \sqrt{\frac{2I(I+1)}{2I+1}}  d^{\Fu}_{\bar{m}',m+1}(-\tu)d^{\Fd }_{m,\bar{m}}(\td) (-1)^{(\Fd-m)}
\begin{pmatrix}
\Fu & 1 & \Fd \\
-(m+1)& 1 & m
\end{pmatrix},
\nonumber\\
\left\langle \Fu,\bar{m}'\right| \hat{\bar{H}}^{0,-(2m)}_\textrm{MW} \left|\Fd,\bar{m} \right\rangle &=&  \mu_\text{B} g_J B_{\textrm{MW},0} \sqrt{\frac{2I(I+1)}{2I+1}}  d^{\Fu}_{\bar{m}',m}(-\tu)d^{\Fd }_{m,\bar{m}}(\td)  (-1)^{(\Fd+m-1)}
\begin{pmatrix}
\Fu & 1 & \Fd \\
-m& 0 & m
\end{pmatrix}  ,
\label{eq:MWCouplingsMWPolarComponents}
\end{eqnarray}
with
\begin{eqnarray}
B_{\textrm{MW},0} &=& \frac{B_{\textrm{MW,z}}e^{-i \phi_z}}{2} , \nonumber \\
B_{\textrm{MW},\pm} &=& \frac{
     \mp B_{\textrm{MW},x}  e^{-i\phi_x}
     + i B_{\textrm{MW},y}  e^{-i\phi_y}   }{2\sqrt{2}} .\nonumber
\end{eqnarray}

\section{\label{app:couplings}Microwave coupling of RF dressed states of $^{87}$Rb}
\label{sec:Appendix3}
In the limit of weak static magnetic fields, the microwave couplings between RF-dressed states are given by Eq.\,(\ref{eq:MWCouplingofRFdressedStates}), which indicates that it is convenient to group the couplings between dressed states according to the polarization of the MW field. 
Taking into account the difference between gyromagnetic factors of the two ground state hyperfine manifolds we obtain the results presented below.

The RF-field is taken to be linearly polarised and perpendicular to the static field $B_\textrm{DC}$. The value $\Delta \bar{m}$ given in the table indicates, for $^{87}$Rb, the value of
$ \bar{m} +  \bar{m}'$ in Eq.~(\ref{eq:ResonantCondition}), such that for nearly equal RF Rabi frequencies $\Omega_{\textrm{RF}}^F $ and $\Omega_{\textrm{RF}}^{F+1}$,
we see an indication of the location of five spectral components within one of the seven groups determined by the index $n$ in Eq.~(\ref{eq:ResonantCondition}).  Following Eq.~(\ref{eq:H_MW_dec}), the superscripts of the label $\hat H^{\sigma,n}_{\textrm{MW}}$ indicate the corresponding polarization ($\sigma$) and the shift of the angular frequency of oscillation of the coupling as observed in the dressed frame, i.e.\ $\omega_{\textrm{MW}} + n \omega_{\textrm{RF}}$. With this, the couplings with $n>0$ ($n<0$) lead to resonances red (blue) detuned with respect to the hyperfine splitting.

The tables below display the coupling between RF-dressed states normalised to the factor $\hbar \Omega_\sigma =  \frac{1}{16}\sqrt{\frac{3}{2}} |\eta_\sigma|\mu_\text{B} g_J B_{\textrm{MW},\sigma}$, for the $\pi$ and $\sigma_-$ polar components of the MW field. The couplings associated with the $\sigma_+$ polarization can be obtained using relation Eq.\,(\ref{eq:relations}).  


\subsection*{$\pi$ polarised MW field}
\begin{figure*}[h!]
\centering
\includegraphics[width=\textwidth]{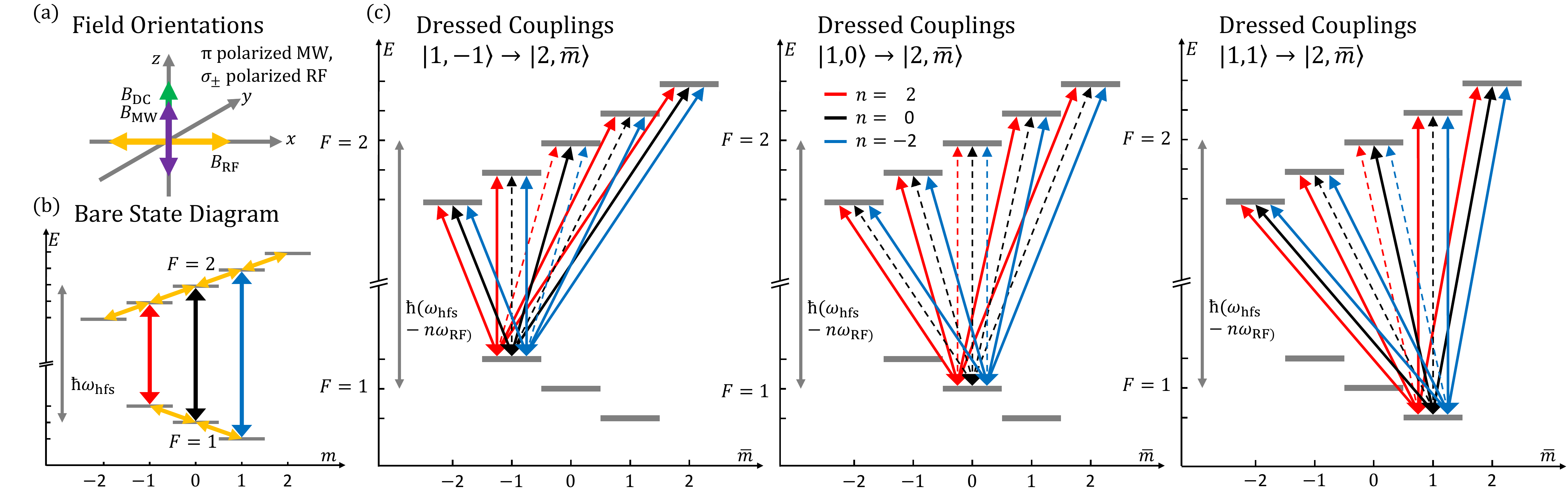}
\caption{\label{fig:MWcouplingPiPolarization}Field configuration with a $\pi$-polarized MW field and sketch of the associated couplings between RF-dressed states. (a) In the lab frame we show the orientation of the DC and AC fields. (b) In the dressed frame we show all the couplings oscillating at the frequency of the MW field $\omega_{\textrm{MW}}$. (c) Couplings between dressed states states $\left|F=1,\bar{m} = 0 \right\rangle \leftrightarrow \left| F=2,\bar{m} \right\rangle$ oscillating at frequencies $\omega_{\textrm{MW}} - 2\omega_{\textrm{RF}}$ (red), $\omega_{\textrm{MW}}$ (black) and $\omega_{\textrm{MW}} + \omega_{\textrm{RF}}$ (blue). In the approximation $g_1=-g_2$, and on RF resonance, some transitions are forbidden, as indicated by dashed lines. The colour code of the amplitude of the couplings is the same as in Table \ref{tb:TablePi}.}
\end{figure*}

\noindent

\begin{table*}[h]
{\scriptsize
\hspace*{-1.5cm}
\begin{tabular}{c|c|c|c|c}
                                             &&                                                                                                               &                                                                                                                                                                                          &                                                                                                                \\
 \large{Coupled}     & \large{$\Delta \bar{m}$}
& \Large{$\frac{H_{\textrm{MW}}^{\pi,n{=}2}}{\hbar \Omega_{0}}$}                                                                                              & \Large{$\frac{H_{\textrm{MW}}^{\pi,n{=}0}}{\hbar \Omega_0}$}                                                                                                                                                                         & \Large{$\frac{H_{\textrm{MW}}^{\pi,n{=}-2}}{\hbar \Omega_0}$}                                                                                              \\
\large{pair}                                       &&                                                                                                               &                                                                                                                                                                                          &                                                                                                                \\
\hline
\hline
$|1,1\rangle \leftrightarrow |2,-2\rangle$  &-1&  \color{red}{$-\sqrt{2}(1-\ctdF)(1+\ctuF)\sin(\tuF)$    }                            & $  -2\sqrt{2}\,\mathrm{sin}\left( \tdF \right) \,{\mathrm{sin}^2\left( \tuF \right) }  $                                                                              &\color{blue}{ $-\sqrt{2}(1+\cos(\tdF))(1-\cos(\tuF))\sin(\tuF)$} \\
$|1,1\rangle \leftrightarrow|2,-1\rangle$ &0& \color{red}{ $-\sqrt{2}(1-\ctdF)(1-\ctuF-2\cos^2(\tuF))$}                                                                               &$ 2\sqrt{2}\,\mathrm{sin}\left( \tdF \right) \,\mathrm{sin}\left( 2\,\tuF \right)   $                                                                      
   & \color{blue}{$\sqrt{2}(1+\ctdF)(1+\ctuF-2\cos^2(\tuF))$ }\\
$|1,1\rangle\leftrightarrow|2,0\rangle$   &1&  \color{red}{$ \sqrt{3} (1-\ctdF)\sin(2\tuF)$      }                                                                & $\frac{4}{\sqrt{3}}\mathrm{sin}\left( \tdF \right) (1-{3 \cos^2\left( \tuF \right) }) $  & \color{blue}{$-\sqrt{3}(1+\ctdF)\sin(2\tuF) $   } \\
$|1,1\rangle\leftrightarrow|2,1\rangle$   &2&  \color{red}{$ \sqrt{2}(1-\ctdF)(1+\ctuF-2\cos^2(\tuF))$  }                                                                         &$  -2\sqrt{2}\,\sin\left( \tdF \right) \,\sin\left( 2\,\tuF \right)   $                                                                                &\color{blue}{ $-\sqrt{2}(1+\ctdF)(1-\ctuF -2\cos^2(\tuF))$  }\\
$|1,1\rangle\leftrightarrow|2,2\rangle$   &3& \color{red}{ $ \sqrt{2}(1-\ctdF)(1-\ctuF)\stuF $         }                                                                            & $  -2\sqrt{2}\,\sin\left( \tdF \right) \,{\sin\left( \tuF \right) }^{2} $                                                                               &\color{blue}{ $\sqrt{2}(1+\ctdF)(1+\ctuF)\stuF $    }         \\

\hline

$|1,0\rangle\leftrightarrow|2,-2\rangle$  &-2& \color{red}{$2 \stdF (1+\ctuF)\stuF$}                                                                                 & $ 4\mathrm{cos}\left( \tdF \right) \,{\mathrm{sin}^2\left( \tuF \right)}$                                                                & \color{blue}{$- 2\stdF(1-\ctuF)\stuF$  }            \\
$|1,0\rangle\leftrightarrow|2,-1\rangle$  &-1& \color{red}{$ 2\stdF (1-\ctuF-2\cos^2(\tuF))$}                                                                         &$  -4\mathrm{cos}\left( \tdF \right) \,\mathrm{sin}\left( 2\,\tuF \right)$                                                                 & \color{blue}{$ 2\stdF(1+\ctuF-2\cos^2(\tuF) )$   }     \\
$|1,0\rangle\leftrightarrow|2,0\rangle$   &0& \color{red}{$ -\sqrt{6}\,\mathrm{sin}\left( \tdF \right) \,\mathrm{sin}\left( 2\,\tuF \right)  $} & $  -4\sqrt{\frac{2}{3}} \mathrm{cos}\left( \tdF \right) (1-{3\mathrm{cos}\left( \tuF \right) }^{2})  $         & \color{blue}{$ - \sqrt{6}\,\mathrm{sin}\left( \tdF \right) \,\mathrm{sin}\left( 2\,\tuF \right)  $ }\\
$|1,0\rangle\leftrightarrow|2,1\rangle$   &1& \color{red}{$ -2\stdF (1+\ctuF-2\cos^2(\tuF)) $     }  & $  4\mathrm{cos}\left( \tdF \right) \,\mathrm{sin}\left( 2\,\tuF \right)   $   & \color{blue}{$ 2\stdF(1+\ctuF+2\cos^2(\tuF))$   }\\
$|1,0\rangle\leftrightarrow|2,2\rangle$   &2& \color{red}{$-2\stdF(1-\ctuF)\stuF $}                                                                                  & $  4\mathrm{cos}\left( \tdF \right) \,{\mathrm{sin}^2\left( \tuF \right) }$                                                               & \color{blue}{$ 2\stdF(1+\ctuF)\stuF$     }                                                                              \\

\hline

$|1,-1\rangle\leftrightarrow|2,-2\rangle$  &-3& \color{red}{ $  -\sqrt{2}(1+\ctdF)(1+\ctuF)\stuF$  }                                                                                   &$ 2\sqrt{2}\,\mathrm{sin}\left( \tdF \right) \,{\mathrm{sin}^2\left( \tuF \right) } $                                                                              & \color{blue}{$  -\sqrt{2}(1-\ctdF)(1-\ctuF)\stuF$    }                         \\
$|1,-1\rangle\leftrightarrow|2,-1\rangle$  &-2&  \color{red}{$ -\sqrt{2} (1+\ctdF)(1-\ctuF-2\cos^2(\tuF))$  }                                                                           & $  -2\sqrt{2}\,\mathrm{sin}\left( \tdF \right) \,\mathrm{sin}\left( 2\,\tuF \right) $                                                                                & \color{blue}{$   \sqrt{2}(1-\ctdF)(1+\ctuF-2\cos^2(\tuF))$       }           \\
$|1,-1\rangle\leftrightarrow|2,0\rangle$   &-1& \color{red}{$  \sqrt{3} (1+\ctdF)\sin(2\tuF)$   }                                                                  & $  -\frac{4}{\sqrt{3}}\mathrm{sin}\left( \tdF \right) (1-{3\mathrm{cos}^2\left( \tuF \right) } )$ &\color{blue}{ $ -\sqrt{3}(1-\ctdF)(\sin(2\tuF))$  } \\
$|1,-1\rangle\leftrightarrow|2,1\rangle$   &0& \color{red}{ $  \sqrt{2}(1+\ctdF)(1+\ctuF-2\cos^2(\tuF))$}                                                                          & $  2\sqrt{2}\mathrm{sin}\left( \tdF \right) \,\mathrm{sin}\left( 2\,\tuF \right)  $                                                                               & \color{blue}{$-\sqrt{2} (1-\ctdF)(1-\ctuF-2\cos^2(\tuF))$    }           \\
$|1,-1\rangle\leftrightarrow|2,2\rangle$   &1&  \color{red}{$  \sqrt{2}(1+\ctdF)(1-\ctuF)\stuF$     }                                                                                 & $  2\sqrt{2}\mathrm{sin}\left( \tdF \right) \,{\mathrm{sin}\left( \tuF \right) }^{2} $   &\color{blue}{ $ \sqrt{2}(1-\ctdF)(1+\ctuF)\stuF$    }         \\
\hline
\end{tabular}
}
\caption{\label{tb:TablePi}Couplings between RF-dressed states induced by a $\pi$-polarised MW field. The colour of the text in the $H_{MW}^\pi/(\hbar\Omega_0)$ columns corresponds to the colours in Figure~\ref{fig:MWcouplingPiPolarization}c, with red, black and blue for groups $n=2, 0, -2$ respectively. The $\sigma_-$ polarization is defined by taking $\phi_z = 0$, $B_{\textrm{MW},z}>0$ and $B_{\textrm{MW},x} = B_{\textrm{MW},y}=0$. }
\end{table*}

\clearpage

\subsection*{$\sigma_-$ polarised MW field}

\begin{figure*}[h]
\centering
\includegraphics[width=\textwidth]{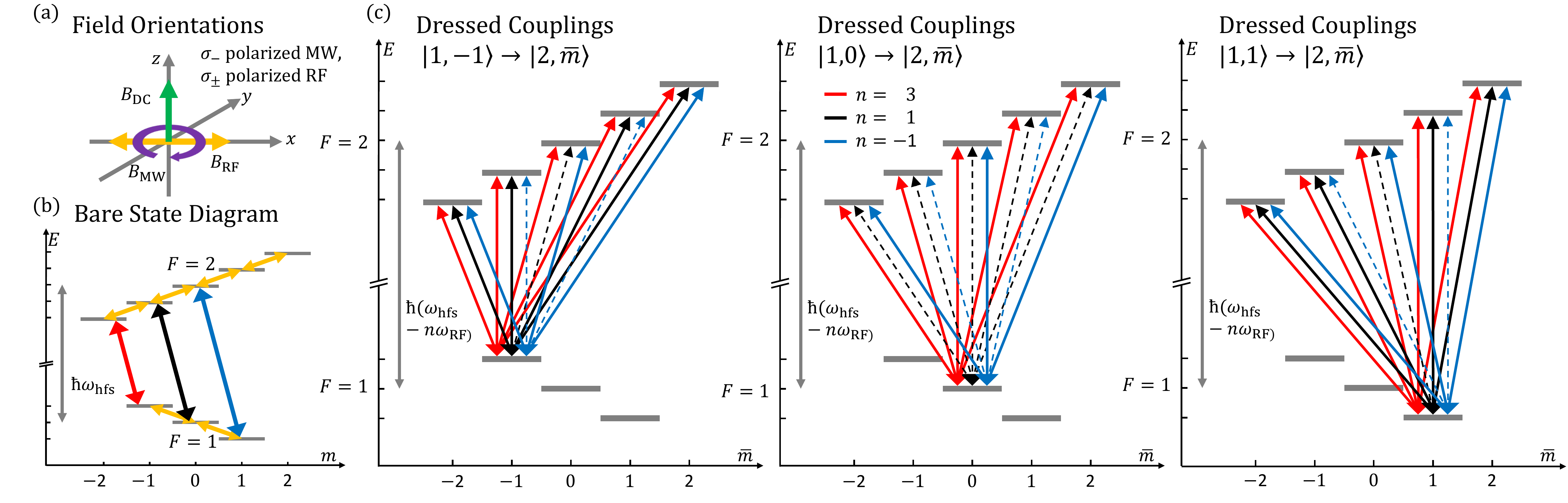}
\caption{\label{fig:MWcouplingSigmaMinusPolarization}Field configuration with a $\sigma_-$-polarized MW field and sketch of the associated couplings between RF-dressed states. (a) In the lab frame we show the orientation of the DC and AC fields. (b) In the dressed frame we show all the couplings oscillating at the frequency of the MW field $\omega_{\textrm{MW}} + \omega_{\textrm{RF}}$. (c) Couplings between dressed states states $\left|F=1,\bar{m} = 0 \right\rangle \leftrightarrow \left| F=2,\bar{m} \right\rangle$ oscillating at frequencies $\omega_{\textrm{MW}} - \omega_{\textrm{RF}}$ (red), $\omega_{\textrm{MW} + \omega_{\textrm{RF}}}$ (black) and $\omega_{\textrm{MW}} + 3 \omega_{\textrm{RF}}$ (blue). In the approximation $g_1=-g_2$ and on RF resonance some transitions are forbidden, as indicated by dashed lines. The colour code of the amplitude of the couplings is the same as in Table \ref{tb:TableSigmaMinus}. }
\end{figure*}

\begin{table*}[h]
{\scriptsize
\hspace*{-1.5cm}
\begin{tabular}{c|c|c|c|c}
                                                &&                                          &                                       &                                                  \\
 \large{Coupled}                   &\large{$\Delta \bar{m}$} 
 & \Large{$\frac{\bar{H}_{\textrm{MW}}^{-,n{=}3}}{\hbar \Omega_{-1}}$}                     
 & \Large{$\frac{\bar{H}_{\textrm{MW}}^{-,n{=}1}}{\hbar \Omega_{-1}}$}          
 & \Large{$\frac{\bar{H}_{\textrm{MW}}^{-,n{=}-1}}{\hbar \Omega_{-1}}$}                            \\
 \large{pairs}                                          &&                                          &                                       &                                                  \\
\hline
\hline
 $|1, \ 1\rangle\leftrightarrow |2,-2\rangle$   &-1& \color{red}{$(1-\ctdF)(1+\ctuF)^2$     }           & $ 2\stdF(1+\ctuF)\stuF$                 & \color{blue}{$ (1+\ctdF)\sin^2(\tuF)$     }                      \\
 $|1, \ 1\rangle\leftrightarrow |2,-1\rangle$   &0& \color{red}{ $2(1-\ctdF)(1+\ctuF)\stuF $        }        & $ 2\stdF(1-\ctuF-2\cos^2(\tuF))$         & \color{blue}{$ -(1+\ctdF)\sin(2\tuF)$  }                         \\
 $|1, \ 1\rangle\leftrightarrow |2, \ 0\rangle$ &1& \color{red}{ $\sqrt{6}(1-\ctdF)\sin^2(\tuF) $ }& $ -\sqrt{6}\stdF\sin(2\tuF)$             &\color{blue}{ $ -\sqrt{\frac{2}{3}}(1+\ctdF)(1-3 \cos^2(\tuF) )$ }\\
 $|1, \ 1\rangle\leftrightarrow |2, \ 1\rangle$ &2& \color{red}{ $2(1-\ctdF)(1-\ctuF)\stuF$           }       &$ -2\stdF(1+\ctuF-2\cos^2(\tuF))$  & \color{blue}{$ (1+\ctdF)\sin(2\tuF)$ }                           \\
 $|1, \ 1\rangle\leftrightarrow |2, \ 2\rangle$ &3&  \color{red}{$(1-\ctdF)(1-\ctuF)^2$ }               & $ -2\stdF (1-\ctuF)\stuF$              &\color{blue}{ $ (1+\ctdF)\sin^2(\tuF)$   }                        \\
\hline

$|1, \ 0\rangle\leftrightarrow|2,-2\rangle$     &-2& 
\color{red}{$-\sqrt{2}\stdF (1+\ctuF)^2$}                & $-2\sqrt{2}\ctdF(1+\ctuF)\stuF $         &\color{blue}{ $\sqrt{2}\stdF \sin^2(\tuF) $                  }\\
$|1, \ 0\rangle\leftrightarrow|2,-1\rangle$     &-1& 
\color{red}{$-2\sqrt{2}\stdF (1+\ctuF)\stuF $     }   & $-2\sqrt{2}\ctdF(1-\ctuF - 2\cos^2(\tuF)) $ & \color{blue}{$-\sqrt{2}\stdF \sin^2(2\tuF) $ }      \\
$|1, \ 0\rangle\leftrightarrow|2, \ 0\rangle$   &0& 
\color{red}{$-2\sqrt{3}\stdF \sin^2(\tuF)$}              & $2\sqrt{3}\ctdF \sin(2\tuF)$            & \color{blue}{$-\sqrt{\frac{4}{3}}\stdF(1-3\cos^2(\tuF))$ }   \\
$|1, \ 0\rangle\leftrightarrow|2, \ 1\rangle$   &1& 
\color{red}{$-2\sqrt{2}\stdF (1-\ctuF)\stuF$ }           &$2\sqrt{2}\ctdF (1+\ctuF-2\cos^2(\tuF))$ & \color{blue}{$\sqrt{2}\stdF \,\sin(2\tuF) $   }               \\
$|1, \ 0\rangle\leftrightarrow|2, \ 2\rangle$   &2& 
\color{red}{$-\sqrt{2}\stdF (1-\ctuF)^2$}     &$2\sqrt{2}\ctdF (1-\ctuF)\stuF$          & \color{blue}{$\sqrt{2}\stdF \,\sin^2(\tuF)$  }                \\
\hline

$|1,-1\rangle\leftrightarrow|2,-2\rangle$       &-3&  \color{red}{$(1+\ctdF)(1+\ctuF)^2$          }          &$-2\stdF (1+\ctuF)\stuF$                 &\color{blue}{ $\left( 1-\ctdF \right) \,\sin^2(\tuF)$   }         \\
$|1,-1\rangle\leftrightarrow|2,-1\rangle$       &-2& \color{red}{ $2(1+\ctdF)(1+\ctuF)\stuF$  }                & $-2\stdF (1-\ctuF-2\cos^2(\tuF))$       & \color{blue}{$-\left(1- \ctdF \right) \,\sin(2\tuF) $  }         \\
$|1,-1\rangle\leftrightarrow|2, \ 0\rangle$     &-1&  \color{red}{$\sqrt{6}(1+\ctdF)\sin^2(\tuF) $} & $\sqrt{6}\stdF\sin(2\tuF)$             &\color{blue}{ $ -\sqrt{\frac{2}{3}}(1-\ctdF)(1-3\cos^2(\tuF))$}   \\
$|1,-1\rangle\leftrightarrow|2, \ 1\rangle$     &0& \color{red}{ $2(1+\ctdF)(1-\ctuF)\stuF $}                 & $ 2\stdF (1+\ctuF-2\cos^2(\tuF))$        & \color{blue}{$\left( 1-\ctdF \right) \,\sin(2\tuF) $      }      \\
$|1,-1\rangle\leftrightarrow|2, \ 2\rangle$     &1& \color{red}{ $(1+\ctdF)(1-\ctuF)^2 $    }              & $ 2\stdF (1-\ctuF)\stuF$                &\color{blue}{ $\left( 1-\ctdF \right) \,\sin^2(\tuF)$    }        \\ 
\hline
\end{tabular}
}
\caption{\label{tb:TableSigmaMinus}Couplings between RF-dressed states induced by a $\sigma_-$-polarised MW field. The colour of the text in the $H_{MW}^-/(\hbar\Omega_{-1})$ columns corresponds to those found in Figure~\ref{fig:MWcouplingSigmaMinusPolarization}c, with red, black and blue for groups $n=3, 1, -1$ respectively. The $\sigma_-$ polarization is defined by taking $\phi_x = 0$, $\phi_y=\pi/2$, $B_{\textrm{MW},z}=0$ and $B_{\textrm{MW},x} = B_{\textrm{MW},y}>0$.}
\end{table*}

\end{widetext}
\bibliographystyle{apsrev4-1}
\bibliography{main}

\end{document}